\documentclass[a4paper,notitlepage,longbibliography]{revtex4-1}
\pdfoutput=1
\usepackage{amsmath}
\usepackage{graphicx}
\usepackage{dcolumn}
\usepackage{multirow}
\usepackage{booktabs}
\newcolumntype{d}[1]{D{.}{.}{#1}}

\begin{document}

\title{High-energy pulse self-compression and ultraviolet generation through soliton dynamics in hollow capillary fibres\footnote{Published in: Nature Photonics \textbf{13}, 8, 547--554, (2019). https://doi.org/10.1038/s41566-019-0416-4}}

\author{John C. Travers}
\email{j.travers@hw.ac.uk}
\homepage{http://lupo-lab.com}
\author{Teodora F. Grigorova}
\author{Christian Brahms}
\author{Federico Belli}
\affiliation{School of Engineering and Physical Sciences, Heriot-Watt University, Edinburgh, EH14 4AS, United Kingdom}

\begin{abstract}
Optical soliton dynamics can cause the extreme alteration of the temporal and spectral shape of a propagating light pulse.
They occur at up to kilowatt peak powers in glass-core optical fibres and the gigawatt level in gas-filled microstructured hollow-core fibres.
Here we demonstrate optical soliton dynamics in large-core hollow capillary fibres.
This enables scaling of soliton effects by several orders of magnitude to the multi-mJ energy and terawatt peak power level.
We experimentally demonstrate two key soliton effects.
First, we observe self-compression to sub-cycle pulses and infer the creation of sub-femtosecond field waveforms---a route to high-power optical attosecond pulse generation.
Second, we efficiently generate continuously tunable high-energy (1--16~$\mu$J) pulses in the vacuum and deep ultraviolet (110~nm to 400~nm) through resonant dispersive-wave emission.
These results promise to be the foundation of a new generation of table-top light sources for ultrafast strong-field physics and advanced spectroscopy.

\end{abstract}

\maketitle

Hollow capillary fibres (HCF)---consisting of a simple circular bore in a glass fibre---have been used in nonlinear optics since the 1970s~\cite{Ippen1970,Miles1977}. Among many applications, they have been used for frequency conversion and amplification through four-wave mixing~\cite{Durfee1997,Kida2014}, vacuum ultraviolet (VUV) generation~\cite{Misoguti2001}, spatio-temporal self-compression~\cite{Wagner2004,Anderson2014,Gao2018}, high-harmonic generation~\cite{Durfee1999,Popmintchev2012} and nonlinear optics in liquids~\cite{chemnitz2017hybrid}. In ultrafast optics their primary application is to pulse compression based on nonlinear spectral broadening~\cite{Nisoli1996,Nisoli1998}. Notably, there is one very important class of nonlinear effect that has not been demonstrated in gas-filled HCF: optical solitons and the rich dynamics associated with them.

Bright temporal optical solitons are obtained by balancing a positive nonlinear refractive index with negative (anomalous) linear dispersion~\cite{Shabat1972,Hasegawa1973}. They underlie a wide range of important phenomena in nonlinear optics, especially in fibres. Some key examples are extreme pulse self-compression~\cite{Mollenauer1983,Im2010b,Joly2011,Mak2013,Balciunas2015}, the formation of white-light supercontinua~\cite{Dudley2006}, and the efficient generation of frequency-tunable pulses through resonant dispersive-wave (RDW) emission~\cite{Wai1986,Im2010b,Joly2011,Mak2013b,Belli2015,Ermolov2015,Bromberger2015,Kottig2017}.
These effects have been thought impossible to achieve in HCF in the visible and near-infrared (NIR)~\cite{Travers2011,Russell2014} due to their weak dispersion in these spectral regions.

The group velocity dispersion (GVD) of the HE$_{nm}$ modes of a gas-filled hollow fibre is given by
\begin{equation}
\label{eqn:b2}
\beta_2(\lambda) \approx \frac{\lambda^3}{4\pi c^2}\left (\rho_r\frac{\partial^2\chi_e}{\partial\lambda^2}-\frac{u_{nm}^2}{2\pi^2a^2}\right ),
\end{equation}
where $n$ and $m$ are the mode indices, $u_{nm}$ is the m$^\mathrm{th}$ zero of the Bessel function $J_{n-1}$, $c$ is the speed of light, $a$ is the HCF core radius, $\lambda$ is the wavelength, $\rho_r$ is the gas density relative to some standard conditions, and $\chi_e(\lambda)$ is the linear electric susceptibility of the gas at those standard conditions (see  supplementary text).
As the first term in Eq.~\ref{eqn:b2} is positive at optical wavelengths for nearly all gases, negative GVD can only be obtained using the second term, which describes the dispersion of the waveguide. Importantly, this waveguide contribution is weaker for larger core sizes.

Hollow-core photonic-crystal fibre (HC-PCF) can offer low-loss guidance despite their small core radius (typically around 15~$\mu$m), leading to sufficient anomalous waveguide dispersion for soliton effects to be observed at low pulse energy~\cite{Ouzounov2003,Travers2011,Russell2014,Markos2017,Saleh2016}.
In comparison, the loss of a conventional HCF with such a small core is prohibitively high~\cite{Marcatili1964}.
The $1/\mathrm{e}$ power loss length is given by
\begin{equation}
\label{eqn:loss}
L_\mathrm{loss}\approx \frac{4\pi^2a^3}{3u_{nm}^2\lambda^2},
\end{equation}
which is just 1~cm for the HE$_{11}$ mode in a 15~$\mu$m radius HCF at a wavelength of 800~nm.

\begin{figure}[h!tb]
    \centering
    \includegraphics{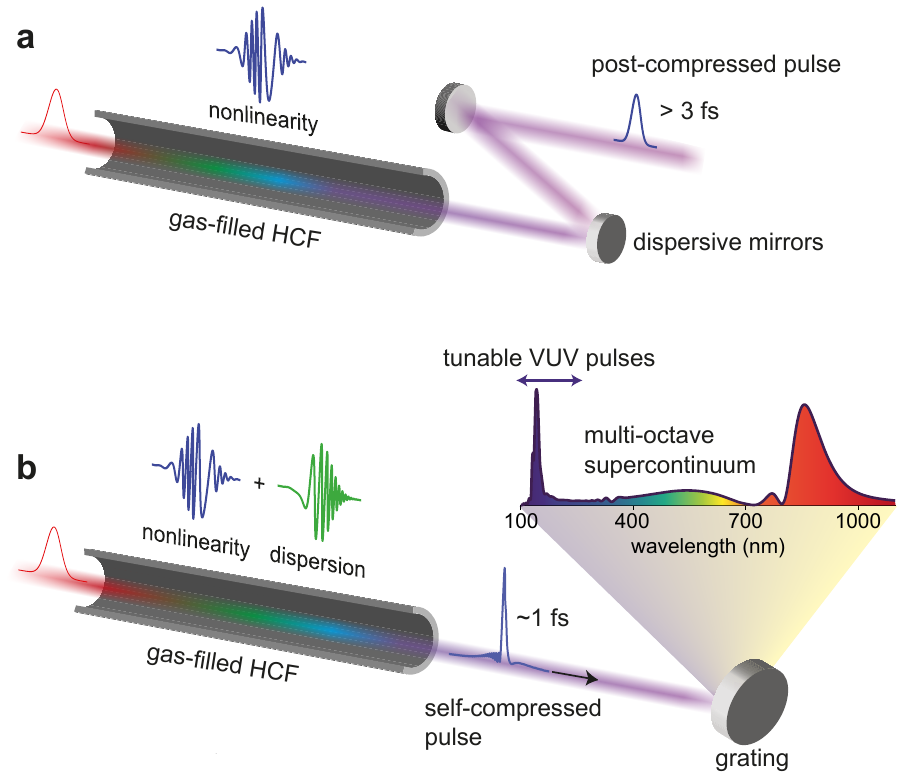}
    \caption{\label{fig:hcf} \textbf{Comparison between post-compression and soliton dynamics in HCF.}
        \textbf{a} Conventional post-compression in gas-filled HCF, where the dynamics in the HCF are dominated by nonlinear spectral broadening based on self-phase modulation and self-steepening, and the anomalous dispersion for pulse compression is provided by chirped mirrors.
        \textbf{b} Soliton dynamics in gas-filled HCF, where the dynamics in the HCF combine both nonlinearity and dispersion, leading directly to extreme self-compression, UV emission and multi-octave supercontinuum formation.}
\end{figure}

The use of HCF with much larger core radii (around 75--500~$\mu$m) reduces the loss dramatically ($L_\mathrm{loss}\propto a^3$) and also allows for much higher pulse energy and peak power than in HC-PCF.
However, it strongly reduces the waveguide dispersion, and so the gas nonlinearity dominates.
In conventional HCF pulse compression, propagation in the hollow fibre is used only to increase the pulse bandwidth through nonlinear self-phase modulation (SPM) and self-steepening, and dispersion in the HCF is neglected.
Instead, linear dispersive mirrors after the fibre provide the negative quadratic phase required to temporally compress the pulse (we refer to this as post-compression).
This is shown schematically in Fig.~\ref{fig:hcf}a.
In this way compression to few- and even single-cycle pulses can be achieved at high pulse energies up to a few mJ~\cite{Nisoli1998,Robinson2006,Bohman2010,Bohle2014,Cardin2015,Silva2018,Jeong2018}.

Theoretical suggestions of how to obtain soliton effects in HCF---which require nonlinearity and dispersion to act simultaneously within the fibre---have included the use of metal-coated HCF to reduce the loss~\cite{Husakou2009}, pumping in higher-order modes to increase $u_{nm}$ in Eq.~\ref{eqn:b2}~\cite{LopezZubieta2018,LopezZubieta2018b}, or moving to longer wavelengths~\cite{Voronin2014,Zhao2017}. However, only numerical results on self-compression have been reported so far, with the rich variety of soliton-induced effects remaining unexplored, and no experiments performed to date.

In this paper we determine the general scaling rules for soliton dynamics in HCF and find that they can in fact be broadly accessed in the fundamental mode and without any coatings or microstructure, even in the visible and NIR. This is possible simply by using shorter pump pulses, longer HCF lengths than usual---enabled by the important innovation of HCF stretching~\cite{Nagy2008,Nagy2011,Bohle2014,Cardin2015,Jeong2018}---or both of these approaches simultaneously.
We experimentally demonstrate two key effects, illustrated in Fig.~\ref{fig:hcf}b.
We observe soliton self-compression to sub-cycle pulses with 43~GW peak power, and soliton-driven ultraviolet (UV) pulse generation between 110~nm and $>350$~nm with energies of 1 to 16~$\mu$J---the highest energies ever demonstrated in this spectral region from a tunable source of ultrafast pulses.
These results represent a power and energy scaling of up to 1000 times compared to previous results in HC-PCF.
Our numerical simulations---extensively verified by experiment---predict that these dynamics can be further scaled to achieve self-compression at the terawatt level, providing a route to optical attosecond pulses~\cite{Hassan2016} with unprecedented power, as well as table-top generation of tunable and ultrafast UV--VUV (3--12~eV) pulses at energies over $100$~$\mu$J, with a peak brightness exceeding free-electron laser systems.

\section{Results}
\subsection{Scaling rules for soliton dynamics in HCF}
The character of ultrashort pulse propagation depends on the full wavelength-dependent dispersion landscape and the balance between nonlinearity and dispersion. This balance is characterised by the pump soliton order $N=(L_\mathrm{d}/L_\mathrm{nl})^{1/2}$, where the dispersion length $L_\mathrm{d}$ and the nonlinear length $L_\mathrm{nl}$ describe the characteristic length scales of GVD and SPM, respectively~\cite{Agrawal2007}. In a hollow core fibre a reasonable parametrisation of the dispersion landscape is the relative location of the pump wavelength $\lambda_0$ with respect to the zero dispersion wavelength $\lambda_\mathrm{zd}$ (defined via $\beta_2(\lambda_\mathrm{zd}) = 0$). This determines the phase-matched location for RDW generation and therefore, in the usual case in which $\lambda_0$ is fixed, changing $\lambda_\mathrm{zd}$ (via the gas pressure) tunes the RDW emission wavelength. It also determines the degree of high-quality self-compression which can be achieved. For the strongest unperturbed compression we require a broad anomalous dispersion region, with $\lambda_\mathrm{zd}$ tuned far from $\lambda_0$.

\begin{figure}[h!tb]
    \centering
    \includegraphics{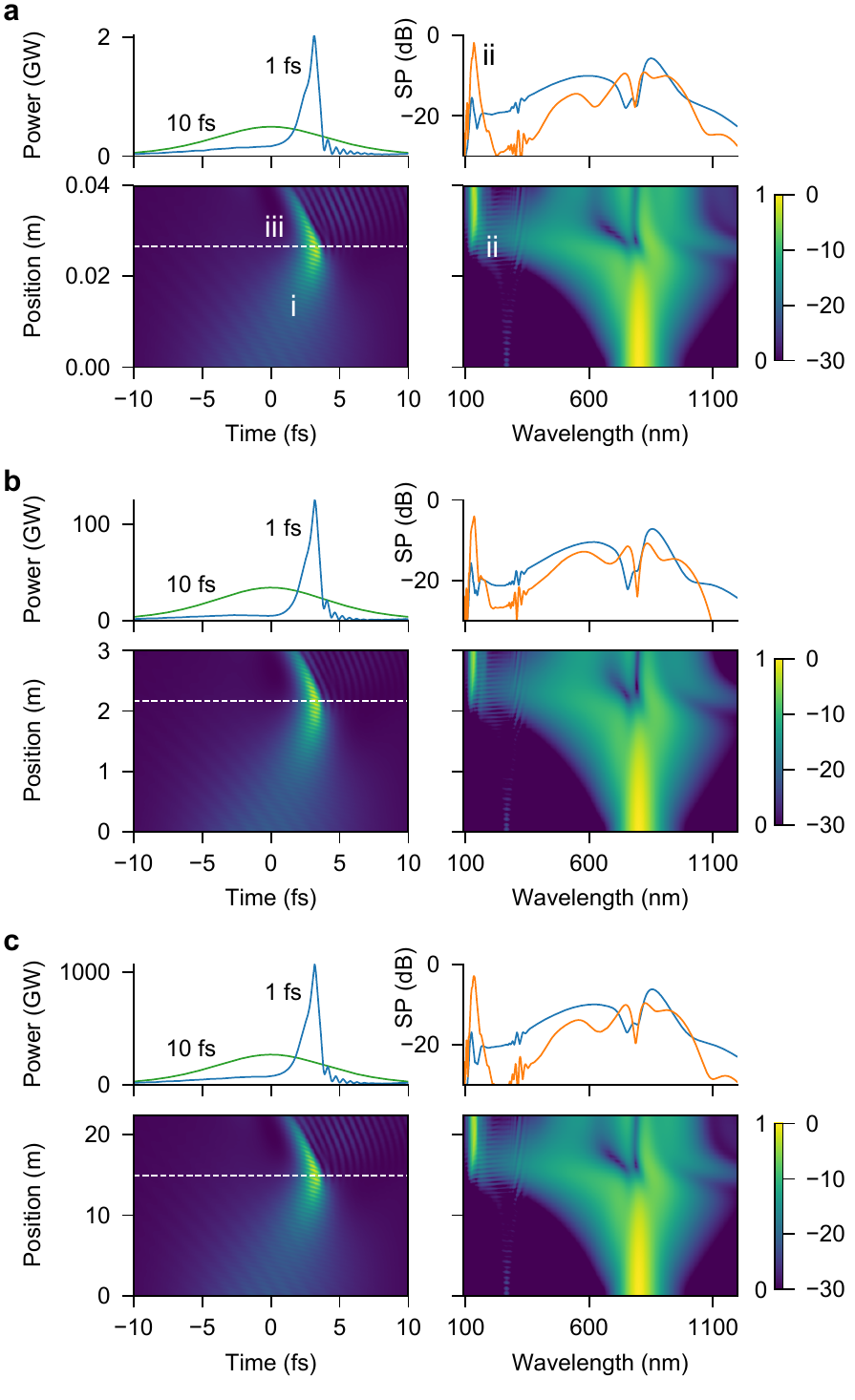}
    \caption{\label{fig:prop} \textbf{Scaling of soliton dynamics in gas-filled hollow fibres.}
        Numerically modelled propagation of 10~fs, $N=2.6$ solitons at $\lambda_0=800$~nm, in hollow fibres filled with helium such that $\lambda_\mathrm{zd}=380$~nm.
        The colour density plots show the evolution of the temporal (linear scale) and spectral power (dB scale).
        The time-domain line plots show the input pulse (green) and the maximally self-compressed pulse (blue), which occurs at the position marked with the white dashed lines.
        The blue spectral line plot shows the corresponding supercontinuum spectrum and the orange spectral line plot shows the spectrum at the end of the propagation. \textbf{a} Example of dynamics in a small core, similar to HC-PCF (but neglecting resonances) with no loss: $a=15$~$\mu$m, 27.8~bar helium, and an energy of 0.006~mJ.
        \textbf{b} HCF with $a=125$~$\mu$m, 400~mb helium, and an energy of 0.4~mJ.
        \textbf{c} HCF with $a=350$~$\mu$m, 51~mb helium, and an energy of 3~mJ. SP is relative spectral power density.
        See supplementary figure Fig.~S1 for a spectrogram depiction of these dynamics.}
\end{figure}

Figure~\ref{fig:prop} shows numerically modelled propagation of 10~fs pulses at 800~nm in helium-filled hollow fibres with very different core sizes (see Methods). For each core size we tune the gas pressure so that $\lambda_\mathrm{zd}=380$~nm (matching the self-compression experiments described later), and adjust the energy to maintain the soliton order at $N=2.6$. Consequently we obtain nearly identical temporal and spectral dynamics even though the energy is increased by almost three orders of magnitude (0.006~mJ to 3~mJ). The small differences---apart from the length scale over which the dynamics occur---are due to different propagation losses in each case.
This scaling is a manifestation of the general scaling of nonlinear optics in gases~\cite{Heyl2016}, which---when applied to hollow fibres---states that if one scales the core radius by a factor $\eta$, exactly the same dynamics (excluding losses) occur if one also scales the gas pressure by $1/\eta^2$ and the fibre length and peak power by $\eta^2$.

The low-energy results in Fig.~\ref{fig:prop}a correspond to an idealised small-core anti-resonant guiding HC-PCF without any loss. In real HC-PCF, cladding resonances inherent to their guidance mechanism would create multiple bands of large loss and dispersion~\cite{Tani2018}, especially in the VUV part of the spectrum. Nevertheless, dynamics similar to those in Fig.~\ref{fig:prop}a have been previously demonstrated~\cite{Im2010b,Joly2011,Mak2013,Balciunas2015,Mak2013b,Belli2015,Ermolov2015,Kottig2017} and utilised in both angle-resolved photoemission spectroscopy \cite{Bromberger2015}, and time-resolved photoelectron imaging spectroscopy~\cite{Kotsina2019}. Fig.~\ref{fig:prop}b and c show results for simple, non-microstructured, HCF. The larger core sizes in HCF enable significant energy scaling.

Self-compression is observable in Fig.~\ref{fig:prop} (marked i) because the pump pulse parameters correspond to a high-order soliton ($N>1$) propagating in the negative dispersion region. Therefore, SPM initially dominates, broadening the pulse spectrum; this enhances the effect of the negative GVD, which compensates for the nonlinear phase and results in temporal compression of the pulse~\cite{Shabat1972,Mollenauer1983,Dudley2006,Agrawal2007} and subsequent soliton fission dynamics~\cite{Dudley2006,Beaud1987,Kodama1987,Husakou2001} (see supplementary figure Fig.~S1 for a spectrogram depiction of these dynamics). In the line plots in Fig.~\ref{fig:prop} we see that this self-compression produces 1~fs pulses. For the small-core case this 1~fs pulse reaches a peak power of 2~GW, whereas in the largest core it is scaled to 1.2~TW.

In the spectral domain, the self-compressed pulse corresponds to a supercontinuum spanning four octaves, starting from 100~nm (blue spectral line plots).
The subsequent spectral dynamics show efficient RDW emission around 130~nm for all core sizes (marked ii), with VUV energies of 0.6~$\mu$J, 25~$\mu$J and 300~$\mu$J for Fig.~\ref{fig:prop} a, b and c respectively. We extract the VUV pulses by spectral filtering (see supplementary figure Fig.~S2). They reach maximum peak powers of 0.2~GW, 6~GW and 80~GW, with a pulse duration of 1.8~fs and a bandwidth limit of 0.8 fs. 

Due to the high intensities reached in Fig.~\ref{fig:prop}---up to $8\times10^{14}$~W/cm$^2$ in all cases---as much as 0.3\% of the gas is ionised (leading to a 40\% energy loss) and plasma effects play a role. In particular, soliton self-frequency blue-shift, along with corresponding soliton acceleration (identified by the curved trajectory of the soliton in the time-domain plots marked iii), are evident~\cite{Holzer2011,Saleh2011}, and these both shift and enhance the strength of the VUV RDW emission~\cite{Saleh2011,Ermolov2015}. These effects are not required for self-compression and RDW emission, both of which still occur with ionisation and plasma effects turned off in the simulations (see supplementary Fig.~S3) or for other sets of parameters which require lower pulse intensities (higher pressures and longer $\lambda_\mathrm{zd}$).

Figure~\ref{fig:prop} is a specific example of soliton dynamics in HCF. In general we would like to vary the pump pulse duration, the zero dispersion wavelength $\lambda_\mathrm{zd}$ (which tunes the RDW emission wavelength~\cite{Joly2011,Mak2013b}) and the core size (to control the energy and power level). To ensure that the soliton dynamics are not precluded by the HCF loss, we need to maintain $L_\mathrm{fiss}<L_\mathrm{loss}$, where $L_\mathrm{fiss}\approx L_\mathrm{d}/N$ is the soliton fission length, which approximates the required length scale for soliton self-compression and subsequent pulse breakup~\cite{Dudley2006}. It can be shown (see supplementary text) that in HCF this length is
\begin{equation}
\label{eqn:lfiss}
L_\mathrm{fiss}\approx \frac{a^2\tau_\mathrm{fw}^2}{{3N\left|\delta (\lambda_0, \lambda_\mathrm{zd})\right|}}\propto \frac{a^2\tau_\mathrm{fw}}{\sqrt{I_0}},
\end{equation}
where $\tau_\mathrm{fw}$ is the full-width pulse duration at half maximum power (FWHM), $I_0$ is the peak intensity, and $\delta (\lambda, \lambda_\mathrm{zd})$ is a function that describes the dispersion of the filling gas.

\begin{figure}[h!tb]
    \centering
    \includegraphics{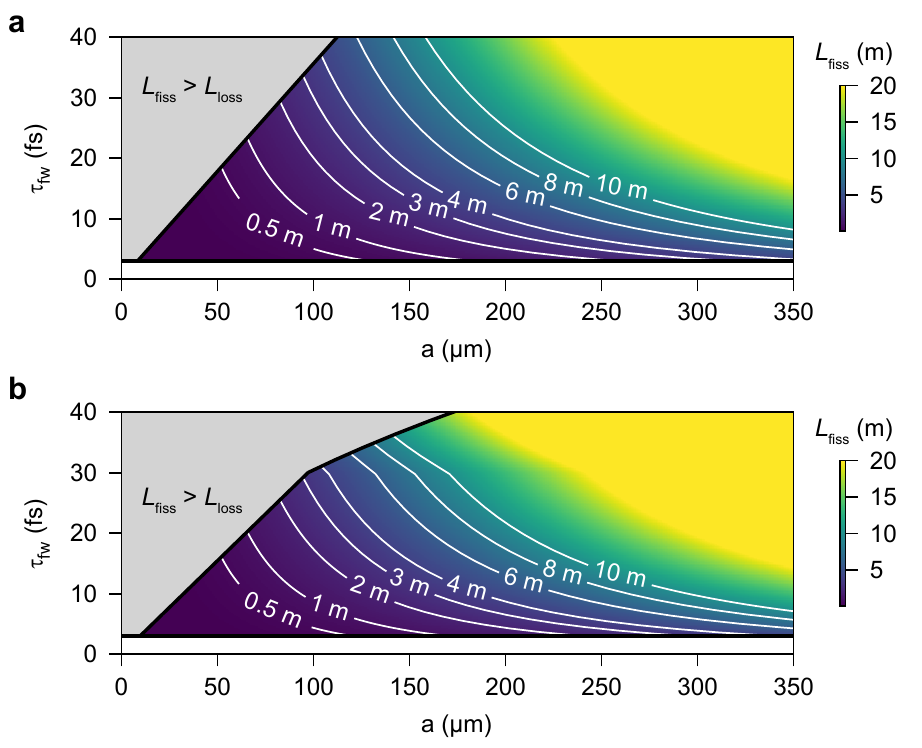}
    \caption{\label{fig:fiss} \textbf{Soliton fission length scaling in helium-filled hollow capillary fibres.} The contours and colour scale show the minimum soliton fission length $L_\mathrm{fiss}$, calculated from Eq.~\ref{eqn:lfiss} using the largest permissible $N$, as a function of core radius $a$ and pulse duration $\tau_{fw}$ for $\lambda_0=800$~nm and \textbf{a}: $\lambda_\mathrm{zd}=380$~nm; and \textbf{b}: $\lambda_\mathrm{zd}=700$~nm. In the grey shaded region it is impossible to obtain $L_\mathrm{fiss}<L_\mathrm{loss}$.}
\end{figure}

Fig.~\ref{fig:fiss} shows this scaling rule for $\lambda_0=800$~nm as a function of core radius and pump pulse duration for two example values of $\lambda_\mathrm{zd}$.
The minimum $L_\mathrm{fiss}$ is obtained for each pulse duration by finding the maximum $N$ or $I_0$, which is determined by several factors.
Firstly, to obtain coherent soliton self-compression as opposed to incoherent modulational instability dynamics we must ensure $N<15$~\cite{Dudley2006}.
Secondly, $N$ increases with the peak power and intensity of the pump pulse, but these are limited by self-focusing~\cite{Fibich2000} and ionisation, respectively (see supplementary text).

The grey regions in Fig.~\ref{fig:fiss}a and b show where it is impossible to observe soliton dynamics because the propagation loss is too high and $L_\mathrm{fiss}>L_\mathrm{loss}$.
This occurs for small core sizes and long pulse durations.
However, $L_\mathrm{fiss}\propto a^2$ whereas $L_\mathrm{loss}\propto a^3$ (Eq.~\ref{eqn:lfiss} and Eq.~\ref{eqn:loss}), so it is always possible to obtain $L_\mathrm{fiss}<L_\mathrm{loss}$ by moving to larger core sizes (right-hand side of Fig.~\ref{fig:fiss}a or b), at the cost of increasing the length of HCF required.
This, in turn, can be remedied by pumping with shorter pulses---an initial duration of 10~fs or less keeps the fission length below 3~m for core radii up to $a=200$~$\mu$m, and even below half a metre for the smallest core sizes.

Until recently, most conventional HCF post-compression systems were limited to HCF lengths of around 1~m.
This is much shorter than the fission length for the 30~fs pump pulses at 800~nm that are typically used, which explains why soliton fission dynamics have not been observed in HCF previously.

\subsection{High-energy soliton self-compression}
\begin{figure}[h!tb]
    \centering
    \includegraphics{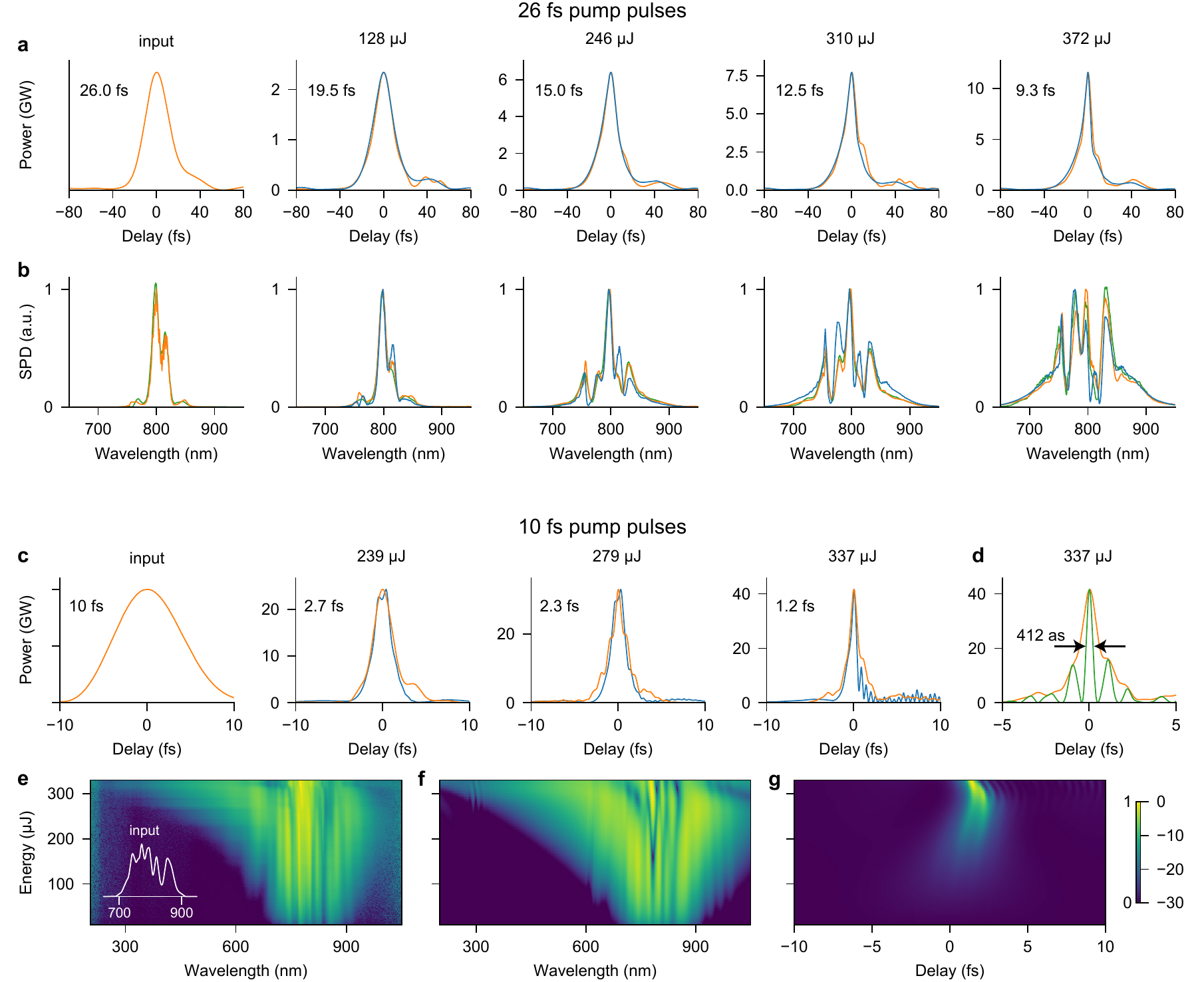}
    \caption{\label{fig:comp} \textbf{Soliton self-compression in gas-filled hollow capillary fibres.}
        \textbf{a} and \textbf{b} are for 26~fs pump pulses and 1~bar helium filling gas.
        \textbf{a} Temporal pulse compression for increasing pump energy; orange: measured from SHG-FROG pulse retrieval (see supplementary Fig.~S4 for details), blue: numerically modelled.
        \textbf{b} Corresponding pulse spectra; orange: FROG retrieved; blue: numerically modelled; green: independently measured.
        \textbf{c}--\textbf{g} Are for 10~fs pump pulses and 0.4~bar helium filling gas.
        \textbf{c} Temporal pulse compression for increasing pump energy; orange: measured from PG-FROG pulse retrieval (see supplementary Fig.~S5 for details) blue: numerically modelled.
        \textbf{d} Closer view of \textbf{c} but with the square of the retrieved electric field shown in green.
        Note the limitations of the measurements underlying the pulse shapes shown in \textbf{c} and \textbf{d} as described in the main text.
        \textbf{e} Measured optical spectrum for increasing pump energy. The inset shows the pump spectrum on a linear scale.
        \textbf{f} Corresponding numerically modelled spectrum.
        \textbf{g} Corresponding numerically modelled temporal pulse shape.
        \textbf{e} and \textbf{f} are on a 30~dB log colour scale, \textbf{g} is on a linear colour scale.
        All results are for a 3~m long HCF with 125~$\mu$m core radius.}
\end{figure}

As an initial experiment to demonstrate self-compression in HCF we used the pulses directly produced by our ti:sapphire system (see Methods) which have 26~fs duration.
We used a 3~m long stretched HCF with $125$~$\mu$m core radius.
At a maximum energy of 400~$\mu$J coupled into the HCF filled with 1~bar of helium ($\lambda_\mathrm{zd} = 470$~nm), the soliton order is $N=6.8$ and the corresponding fission length is 4.2~m.
Although the HCF is too short to reach the fission point, Fig.~\ref{fig:comp}a shows clear experimental evidence of self-compression from 26~fs to 9~fs as we increase the pump energy (and hence the soliton order).
Furthermore, our numerical modelling reproduces the experimental data extremely closely, both in terms of the power spectrum (Fig.~\ref{fig:comp}b) and the fine details of the pulse shape in the time domain (Fig.~\ref{fig:comp}a).
Such good agreement is rarely achieved, and it supports both the accuracy and validity of our numerical model and the fidelity of the pulse measurement.

To observe complete soliton self-compression we start with 10~fs pump pulses produced in a conventional HCF post-compression setup (see Methods).
Figure~\ref{fig:comp}(c--e) shows the experimental results.
At a helium pressure of 0.4~bar ($\lambda_\mathrm{zd}=380$~nm) and an energy of 239~$\mu$J ($N=2$) we obtain self-compression from 10~fs to 2.7~fs, which is close to a single cycle and slightly broader than the bandwidth-limited duration of 2.5~fs.
The pulse becomes even shorter as we increase the energy, compressing to 2.3~fs (bandwidth limit 2.1~fs) at 279~$\mu$J and finally, at 337~$\mu$J, to 1.2~fs for a bandwidth limit of 1.1~fs.
At this point, the full width of the square-field profile, shown in Fig.~\ref{fig:comp}d, is only 412~as---the soliton has self-compressed to an optical attosecond pulse~\cite{Hassan2016}.

These pulse measurements are based on a polarisation-gating frequency-resolved optical gating (PG-FROG) technique which covers the band from 200~nm to $>1100$~nm (see Methods), making them the most broadband FROG measurements ever reported.
Nevertheless, it must be noted that the technique has some severe limitations and the sub-cycle pulse duration cannot be reliably established from our PG-FROG measurements alone.
The pulse profiles shown in Figure~\ref{fig:comp} are obtained by measuring the pulses after they have been stretched by propagation through dispersive materials---most notably the exit window of the HCF system and air in the laboratory---and subsequently numerically back-propagating them to the exit of the HCF.
In general this process is rigorous, with the refractive index of air, silica and MgF2 well-characterised. For the 2.7~fs pulse the back-propagated pulse duration changes by less than 5\% within the experimental uncertainty on the propagation distance and window thickness.
However, the extreme bandwidth and extension into the UV of the sub-cycle pulses means that a difference of only 10~$\mu$m in glass thickness is sufficient to stretch the 1.2~fs pulse by a factor of 2.5.
Since PG-FROG always requires a transmissive interaction medium, even placing the apparatus into vacuum would not avoid this issue.
Fully independent experimental evidence will only be provided by much more complicated experiments such as attosecond streaking~\cite{Wirth2011,Hassan2016}.

However, there are two additional observations that clearly point toward the generation of optical attosecond pulses in our experiments.
The first is the consistent and excellent agreement between all of our experimental data and the numerical model.
This is demonstrated in Fig.~\ref{fig:comp}a and b and again for the particular case of sub-cycle self-compression in Fig.~\ref{fig:comp}e and f, where we numerically reproduce the spectral power density, allowing us to infer the temporal dynamics (Fig.~\ref{fig:comp}g).
The simulated and measured pulse profiles shown in Fig.~\ref{fig:comp}c are also in excellent agreement despite the uncertainties of the measurement.
The second observation is that compression to a sub-cycle pulse is required for RDW emission to occur in the VUV wavelength range. This can be understood from the cascaded four-wave-mixing picture of RDW emission~\cite{Erkintalo2012} in which all of the frequencies between the RDW and the pump need to be present simultaneously. That this is the case is clearly apparent from the spectrograms in supplementary figure Fig.~S1. For RDW emission in the VUV we have the simultaneous presence of frequencies spanning up to 3 octaves, which corresponds to a sub-cycle pulse in the time-domain. The correlation between VUV RDW emission and sub-cycle pulse generation has also been found in HC-PCF~\cite{Belli2015,Ermolov2015}. Therefore, our measurements of RDW emission across the VUV (see below) also support the existence of such extreme pulse compression in HCF---in particular, the parameters for the 1.2~fs pulse in Fig.~\ref{fig:comp}c are the same as for the 130~nm RDW shown in Fig.~\ref{fig:uv}a.

\subsection{Tunable vacuum and deep ultraviolet pulse generation}
\begin{figure}[h!tb]
    \centering
    \includegraphics{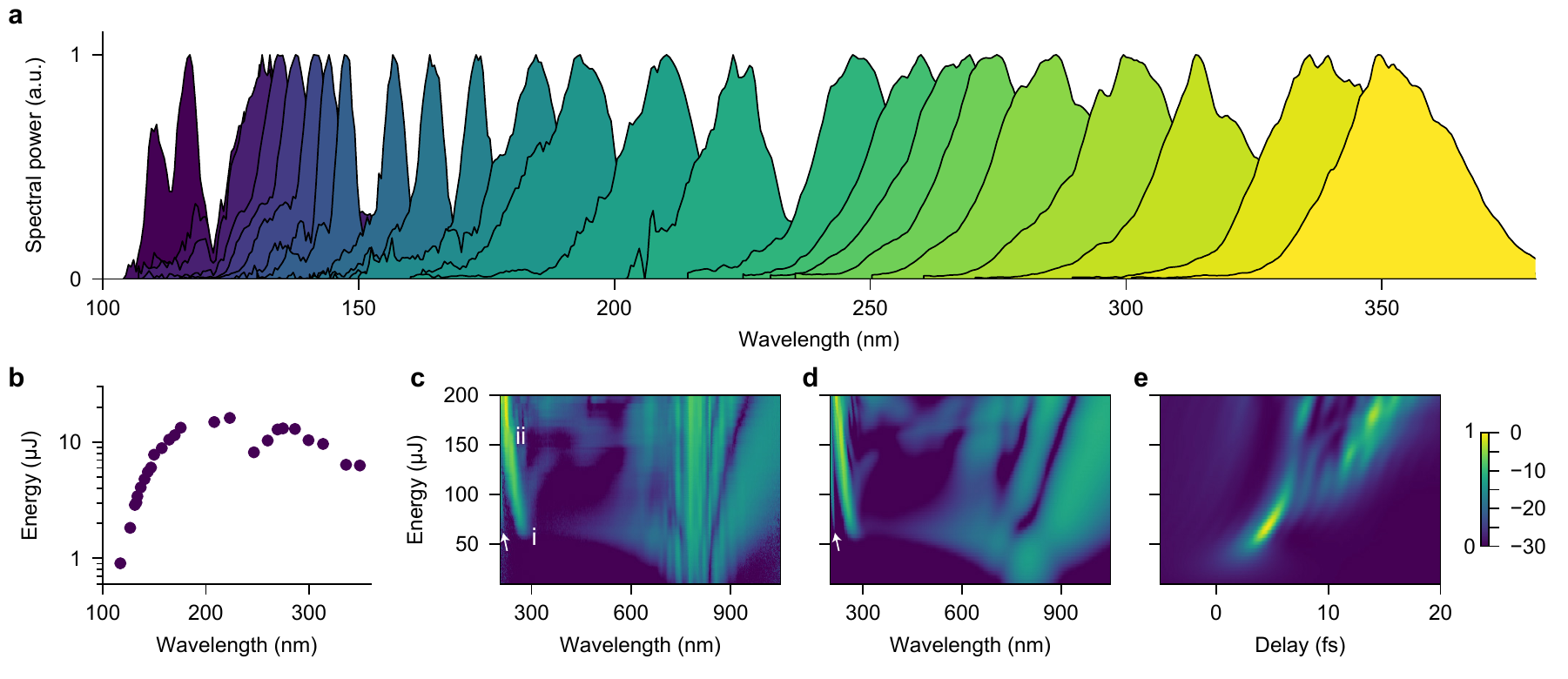}
    \caption{\label{fig:uv} \textbf{Tunable resonant dispersive-wave emission in gas-filled hollow capillary fibres.} \textbf{a} Measured RDW spectra from a 3~m long HCF with 125~$\mu$m core radius filled with helium and pumped with 10~fs pulses (the pump spectrum is shown in the inset to Fig.~\ref{fig:comp}e). The pressures used range from 230~mb (shortest wavelength spectrum) to 4~bar (longest wavelength spectrum). Full parameters are listed in the supplementary material Table~S1. Extension to beyond 400~nm was also measured at higher pressures. \textbf{b} Measured VUV to DUV pulse energies. \textbf{c} Spectral evolution at the output of the HCF for 10~fs pump pulses as a function of input pulse energy for 1~bar neon. \textbf{d} Corresponding numerical simulation. \textbf{e} Corresponding numerical simulation of the temporal power. \textbf{c} and \textbf{d} are on a 30~dB log colour scale; \textbf{e} is on a linear colour scale. The white arrows in \textbf{c} and \textbf{d} indicate the RDW emission in the HE$_{12}$ mode.}
\end{figure}

At the extreme compression point, the soliton breaks up due to higher-order linear and nonlinear effects, such as self-steepening, shock formation and higher-order dispersion~\cite{Dudley2006,Beaud1987,Kodama1987,Husakou2001}.
The latter induces resonant dispersive-wave (RDW) emission---often referred to as emission of fibre-optic Cherenkov radiation or resonant radiation~\cite{Wai1986,Mak2013b,Im2010b,Joly2011,Ermolov2015,Kodama1987,Husakou2001,Akhmediev1995,Erkintalo2012}.
Simulations of the process in HCF are shown in Fig.~\ref{fig:prop}, with RDW emission occurring at the point marked (ii). Further details of the RDW emission dynamics can be seen in the spectrograms in supplementary Fig.~S1. 
The emission wavelength depends on the dispersion landscape and hence on the gas pressure and core size (Eq.~\ref{eqn:b2}).

Fig.~\ref{fig:uv}a shows the experimental tuning of RDW emission from below 110~nm to 350~nm by changing the gas pressure in our 125~$\mu$m HCF pumped with 10~fs pulses (the full parameters are in supplementary Table~S1). Further tuning beyond 400~nm was also obtained at higher pressures. Due to the larger core size, the emitted pulses in the VUV are up to 1000 times more energetic than obtained in HC-PCF, as shown Fig.~\ref{fig:uv}b and supplementary Tables~S1 and S2. Even at the shortest RDW band which spans 107--121~nm the energy is approximately 1~$\mu$J; this increases to around 13~$\mu$J near 180~nm, 16~$\mu$J near 220~nm and remains around 10~$\mu$J until starting to decrease at 300~nm due to the lower pump energy required at higher pressures. The estimated conversion efficiency to the VUV ranges from 0.5\% around 120~nm up to 5\% around 170~nm and in the deep ultraviolet (DUV, 200~nm to 300~nm) this increases to over 15\% (see supplementary Table~S1). These are the highest energies and conversion efficiencies ever produced by a continuously tunable source in this spectral range. Up-scaling of the system as a whole would provide much higher pulse energies at all UV wavelengths (see Fig.~\ref{fig:prop}).

The RDW emission process is highly stable---we measure a relative intensity noise of 1.4\% at the 252~nm RDW peak (see supplementary Fig.~S6). Furthermore, we have not observed any degradation of transmission through the HCF over 15 months of prolonged DUV and VUV generation experiments because the high-intensity ultraviolet light is confined in the hollow core and does not significantly overlap with the glass cladding, whereas the exit window (when used) can be damaged after prolonged exposure to high DUV fluence.

Similar results can be obtained in other gases with the pressure adjusted for their different dispersion and nonlinearity.
As one example, Fig.~\ref{fig:uv}c shows the experimental RDW emission dynamics at 1~bar of neon ($\lambda_\mathrm{zd}=550$~nm).
The RDW wavelength changes from 275~nm (marked i) to 225~nm (marked ii) as the pump energy increases from 60 to 200~$\mu$J (from $N=2.3$ to $N=4.3$).
This is a well-known effect arising from the nonlinear contribution to the phase-matching~\cite{Austin2006,Mak2013b}.
Fig.~\ref{fig:uv}d shows the corresponding numerical simulations, which once again closely reproduce the experiment, including fine details at the $-30$~dB level over the full range of the energy scan.
Fig.~\ref{fig:uv}e shows the corresponding simulations in the time domain, illustrating the initial self-compression process followed by soliton fission.

The RDWs in Fig.~\ref{fig:uv}a and Fig.~\ref{fig:uv}c are in the fundamental mode of the HCF. We have confirmed this by measuring the divergence of the filtered RDW emission at 355~nm and comparing it to the numerically calculated divergence of the fundamental HCF mode. The results are shown in supplementary Fig.~S7, and exhibit almost exact agreement. Weaker RDWs in the HE$_{12}$ mode are also generated and appear as additional features in the spectra (for example, as shown by the white arrows in  Fig.~\ref{fig:uv}c and d), at the corresponding phase-matched point~\cite{Tani2014}.

\section{Discussion}
Advances in the control and measurement of the dynamics of matter with light have been contingent upon our ability to produce light pulses with decreasing temporal duration and wider spectral coverage.
We have shown that soliton dynamics in gas-filled HCF significantly expands our toolset in both of these directions.

While post-compression can be optimised to produce high-quality single-cycle pulses~\cite{Silva2018}, shorter pulses require the use of multiple chirped-mirror sets~\cite{Wirth2011,Hassan2016}.
Soliton self-compression offers a much less complex and lower-cost route to sub-cycle pulse generation.
It can be used in any spectral region, whereas chirped mirrors are designed and manufactured for a specific wavelength range.
By replacing the conventional pre-compressor in our setup with an additional soliton self-compression stage, as discussed in~\cite{Li2010}, the requirement for chirped mirrors could be removed entirely.
Two-stage HCF systems, as used here, have previously been demonstrated for enhanced non-soliton spectral broadening in HCF~\cite{Schenkel2003,Vozzi2005}, and are readily implemented in laboratories with existing HCF post-compressors.
Furthermore, soliton self-compression can be energy-scaled with the fibre core size, the only cost being the increased laboratory length required.
As described in this article, this offers a route to terrawatt-level sub-femtosecond pulse generation, and to optical attosecond pulses 20 times more powerful than state-of-the-art lightfield synthesizers~\cite{Hassan2016}.
The fission length in such a system is around 15~m, as shown in Fig.~\ref{fig:prop}c.
While long, this would be a unique source of 1~fs, 1~TW pulses, and the apparatus would still be smaller than other facility-scale lasers.
As described in Eq.~\ref{eqn:lfiss}, using 5~fs pulses---available from current systems---would halve the fission length to 7.5~m.
That this length-scale is practically feasible has already been demonstrated in conventional compressors using 6~m long HCF~\cite{Jeong2018}.

The ability to scale soliton dynamics by two or more orders of magnitude in HCF goes far beyond pulse compression.
As we have demonstrated, we can use soliton-driven RDW emission as a table-top source of tunable DUV/VUV pulses with high energy and few-cycle duration.
The DUV energies demonstrated here are ten times higher than the best results from HC-PCF~\cite{Kottig2017}, and the VUV energies demonstrated here are up to three orders of magnitude higher~\cite{Ermolov2015}. In both cases the conversion efficiency was higher, and is in fact the highest achieved for RDW emission.
Furthermore, HCF is free of the guidance resonances which are inherent to HC-PCF and which compromise the guidance of UV radiation and reduce the lifetime of the waveguide.
We also achieve two orders of magnitude higher conversion efficiency as compared to previously demonstrated schemes of high-energy VUV generation ~\cite{Misoguti2001,Ghotbi2010,Shi2013,Zhou2014}.
Those schemes were all pumped with much higher energies, achieved similar or lower output energy, could not be tuned as widely (if at all), and produced VUV pulses that were longer than we expect to have produced here (see supplementary Table~S2 for a detailed comparison).
Recent experiments in HC-PCF have shown that RDW emission in the DUV can be extremely short in duration, generating pulses as short as 3~fs~\cite{Brahms2019,Ermolov2016}.
Given the close agreement between our numerical simulations and experiments, we can provide insight into the temporal structure of the VUV RDW (which is beyond the spectral range of our pulse measurement techniques) and confirm the short pulse duration for VUV generation in gas-filled HCF.
For example, the VUV pulse generated in the simulations in Fig.~\ref{fig:prop}b has a duration of just 1.8~fs when it has the highest peak power of 6~GW (see supplementary Fig.~S2). For diffraction limited beams the peak brightness is proportional to peak power (see supplementary text) and so our RDW emission source has a similar peak-brightness to VUV generation in free-electron lasers (FELs), which produce up to gigawatt VUV peak power~\cite{Chang2018,Ayvazyan2002}.
In the predictions for the larger-core HCF (Fig.~\ref{fig:prop}c), the VUV pulse energy is increased to over 300~$\mu$J and the peak power to over 80~GW, thus providing a table-top source with ten times the peak brightness of an FEL in the VUV.

That the full range of soliton dynamics can be accessed in HCF is in itself a remarkable discovery.
Beside sub-cycle self-compression and RDW emission, other phenomena already observed in HC-PCF could also be transferred to HCF, such as Raman-soliton effects~\cite{Belli2015,Belli2018}, soliton-plasma dynamics~\cite{Holzer2011}, among many others~\cite{Travers2011,Russell2014,Markos2017,Saleh2016}.

The new light source of high energy sub-cycle pulses and tunable high-brightness VUV--DUV pulses presented here, along with the rich dynamics from additional soliton effects, yet to be explored in HCF, should form the basis for a new generation of experiments in ultrafast science, advanced time-resolved spectroscopy, and nonlinear optics pumped in the VUV.

\section{Methods}
\subsection{Main experimental setup}
The pump pulses were produced by a commercial ti:sapphire oscillator, regenerative amplifier and single-pass amplifier chain (Coherent Legend Elite Duo USX) producing 26~fs, 800~nm pulses at 1~kHz repetition rate.
The experimental setup is depicted in Fig.~S8.
In a first stage, the 26~fs pump pulses were compressed using an HCF post-compression setup, consisting of a home-built 1.6~m long stretched capillary system~\cite{Nagy2008,Nagy2011}, with 225~$\mu$m inner radius, filled with helium (with a typical throughput at full power, including coupling and transmission, of $>70$\%), followed by 12 reflections from double-angle chirped mirrors (PC70, Ultrafast Innovations), a broadband attenuator based on a $\lambda/2$ waveplate and Brewster-angle reflection from a silicon plate, and anti-reflection coated BK7 wedges for dispersion fine-tuning (Femtolasers).
This system was tuned to compress the pulses at the entrance  of the second-stage HCF.
It was set to either deliver 26~fs (when the first stage HCF was evacuated) or around 10~fs bandwidth-limited pulses when the first-stage helium pressure was 2.2~bar.

The subsequent soliton stage consisted of a home-built 3~m long stretched HCF system with 125~$\mu$m inner radius. The coupling efficiency to this stage was 60--70\%, and the linear transmission is 65\% at 800~nm. The gas pressure could be controlled, and was stable to, within 0.1~mbar.
The windowless output of this stage was directly connected to a vacuum system, containing a fully calibrated, home-built, VUV spectrometer, consisting of either a 1200 G/mm Al+MgF2 coated, or 2400 G/mm Pt coated, concave aberration corrected diffraction grating (McPherson Inc.), two 200~$\mu$m slits, and a VUV silicon diode (SXUV100G, OptoDiode) which has been calibrated at the Physikalisch-Technische Bundesanstalt, Germany (PTB).
Alternatively, the gratings in our VUV spectrometer can be shifted under vacuum, to allow the output of the HCF to pass through our vacuum system, either through a 2~mm silica or 1~mm MgF2 output window, to be measured in atmosphere with: an integrating sphere coupled to a DUV to NIR charge-coupled device (CCD) based spectrometer (calibrated as a system), camera for beam profiling, calibrated thermal power meter, or two different frequency-resolved optical gating (FROG) setups.

\subsection{VUV pulse energy measurements}
The absolute spectral energy measurements are achieved by collecting the entirety of the HCF output beam with the VUV grating and focusing the diffracted beam through a large aperture onto our calibrated diode.
The aperture was set to ensure only the VUV part of the spectrum was collected.
The spectral energy density (in~$\mu$m/nm) was obtained from the photocurrent as follows: the photocurrent was filtered to obtain the zero frequency (DC) component, and processed to account for the pump repetition rate, the absolute diode calibration, the gain of our current amplifier (DLPCA-200, FEMTO Messtechnik GmbH), and the manufacturer supplied theoretical grating efficiency (which provides a conservative lower bound to the estimated energy density, as the actual grating efficiency is always lower than the theoretical one).
Measurements with a silica or BK7 window as a filter were also used to verify the VUV signal below 160 nm and DUV signal below 280~nm, and exclude re-entrant spectra or scatter from the 0$^\mathrm{th}$ order diffraction.

The absolute calibration was verified by independently measuring a RDW generated at 240~nm with 1.8~bar helium with both the VUV spectrometer and our DUV to NIR CCD spectrometer with integrating sphere; the latter was calibrated for absolute spectral irradiance with a PTB traceable deuterium lamp and quartz-tungsten halogen lamp traceable to the United States National Institute of Standards and Technology (NIST). Furthermore, we measured the diffracted DUV signal directly with an additional calibrated photodiode (Ophir PD300-UV). The results all agreed with each other to within 6\%, which is within the stated uncertainty of the calibration sources themselves. The measured energy was 8.5~$\mu$J in agreement with Fig.~\ref{fig:uv}b.

\subsection{Pulse characterisation measurements}
\subsubsection{Second harmonic FROG}
Pulse characterisation of pulses longer than $9$~fs was based on a home-built all-reflective split-mirror second harmonic frequency-resolved optical gating (FROG) device, based on a 10~$\mu$m thick BBO crystal cut for type-I phase-matching.
The raw traces were spectrally corrected using the frequency marginal and retrieved using the extended ptychographic iterative engine (ePIE)~\cite{Sidorenko2016}.
The retrievals were checked to be independent from different initial conditions and recovered the independently measured spectrum (see e.g. Fig.~S4).
The resulting pulses were then back-propagated to account for the air path and 2~mm glass window at the output of the second capillary stage.
The excellent agreement between simulation and pulse measurements (Fig.~\ref{fig:comp}) verifies that this process was reliable.

\subsubsection{Polarisation-gating FROG}
Pulse characterisation of pulses shorter than $9$~fs was carried out using a novel ultrabroadband polarisation-gating frequency-resolved optical gating (PG-FROG) device, the layout of which is shown in Fig.~S9.
The raw traces were corrected for the spectral response of the apparatus as well as the frequency-dependent nonlinearity by using the frequency marginal and the separately measured power spectrum of the unknown pulse.
When taking into account the known spectral sensitivity of the spectrometer and the theoretical frequency dependence of the efficiency of the nonlinear process, the frequency marginal of the trace already closely matches the measured spectrum, however the correction aids the pulse retrieval process.
The raw traces contained an additional signal at the wavelength of the gate pulse caused by scatter in the fused silica sample.
This was removed by high-pass filtering the raw traces in this wavelength region, attenuating delay-independent features.
The pulse retrieval was also carried out using ePIE~\cite{Sidorenko2016}, with the modification that at each iteration, the retrieved gate pulse was updated by numerically back-propagating the retrieved test pulse to the gas cell exit and then forward-propagating it through the optics in the gate pulse arm of the device.
Initial guesses for the test and gate pulses were formed by numerically propagating a transform-limited pulse from the capillary exit to the interaction point, taking into account the air path as well as the optical elements.
This significantly improved both the speed at which the retrieval converges and the retrieval error reached at convergence.
Fig.~S5 shows the measured and retrieved traces.
The resulting pulses were then back-propagated to account for the air path and optics to the output of the second capillary stage.

\subsection{Numerical simulations}
We performed rigorous numerical propagation simulations using our unidirectional, full-field, spatio-temporal propagation code. Our model includes a full vector polarisation model, intermodal coupling, modal dispersion and loss, the Kerr effect and self-focusing as well as photoionisation and plasma dynamics.

Our numerical simulations agree extremely well with experimental data without the use of any adjustable parameters; all material properties and are based on the best available values~\cite{Borzsonyi2008,Ermolov2015,Lehmeier1985}, the stretched HCF is almost exactly modelled by the Marcatili model we employ~\cite{Marcatili1964}, and the input pulse parameters are based purely on our experimental measurements.

We expand the full vectorial spatio-temporal electric field in the frequency domain and over the natural hollow-fibre modes
\begin{equation}
\label{eqn:mode_expansion}
\mathbf{E}\left(t,r,\theta,z\right) = \frac{1}{2\pi}\int_{-\infty}^{+\infty}\mathrm{d}\omega \sum_j \mathbf{\hat{e}}_j(r,\theta)E_j(\omega, z)\exp \left(-i\omega t\right),
\end{equation}
where $r$ and $\theta$ indicate the transverse position in the hollow fibre core, $z$ the axial position in the fibre, $t$ represents time, $\omega$ is angular frequency, $j$ is the mode index, the $\mathbf{\hat{e}}_j$ are the orthonormal transverse electric field shapes of the modes~\cite{Marcatili1964}, and the $E_j$ are the complex spectral amplitudes of the modes which evolve during propagation. Following \cite{Kolesik2004,Tani2014} the spectral amplitudes are coupled via
\begin{equation}
\label{eqn:modeprop}
\partial_zE_j(\omega,z)=\left(i\beta_j(\omega)-i\frac{\omega}{v}-\frac{\alpha_j(\omega)}{2}\right)E_j(\omega,z) + i\frac{\omega}{4}P_j^\mathrm{nl}(\omega,z),
\end{equation}
\begin{equation}
\label{eqn:modeproj}
P_j^\mathrm{nl}(\omega,z) = \int_0^{2\pi}\mathrm{d}\theta\int_0^ar\mathrm{d}r\,\mathbf{\hat{e}}_j^*(r,\theta)\cdot \mathbf{P}^\mathrm{nl}(\omega,r,\theta,z),
\end{equation}
where $\beta_j$ and $\alpha_j$ are the mode propagation and attenuation constants~\cite{Marcatili1964} (see also the supplementary material), $v$ is a suitably chosen, but arbitrary, reference frame velocity, and $\mathbf{P}^\mathrm{nl}(\omega,r,\theta,z)$ is the Fourier transform of the full vector and spatially resolved nonlinear polarisation.

Our model proceeds by numerically integrating Eq.~\ref{eqn:modeprop}. At each step we make use of Eq.~\ref{eqn:mode_expansion} to obtain the full electric field, from which we calculate the full nonlinear polarisation for projection back onto the modes through Eq.~\ref{eqn:modeproj}. This projection is accurately computed using the full two-dimensional integral over the cross-section of the fibre core using a parallel p-adaptive cubature method.

The nonlinear polarisation term, for the purposes of the present article, includes the Kerr effect through
\begin{equation}
\mathbf{P}^\mathrm{Kerr}\left(t,r,\theta,z\right)=\rho_r\epsilon_0\chi^{(3)}\left[\mathbf{E}\left(t,r,\theta,z\right)\cdot \mathbf{E}\left(t,r,\theta,z\right)\right] \mathbf{E}\left(t,r,\theta,z\right),
\end{equation}
where $\epsilon_0$ is the electric permittivity of free space, $\chi^{(3)}$ is the third-order nonlinear susceptibility of the filling gas at some standard conditions~\cite{Lehmeier1985}, and $\rho_r$ is the gas density relative to those conditions.
The ionisation and plasma response is included through~\cite{Geissler1999}
\begin{equation}
\mathbf{P}^\mathrm{ion}\left(t,r,\theta,z\right) = I_p\int_{-\infty}^t \mathrm{d}t'\frac{\partial_t n_e(r,\theta,z)}{\left|\mathbf{E}\left(t',r,\theta,z\right)\right|} + \frac{e^2}{m_e^2}\int_{-\infty}^{t}\mathrm{d}t'\int_{-\infty}^{t'} \mathrm{d}t''n_e(r,\theta,z)\mathbf{E}\left(t'',r,\theta,z\right),
\end{equation}
where the free electron density is given by
\begin{equation}
n_e(r,\theta,z) = n_0\left(1 - \exp \left[-\int_{-\infty}^t\mathrm{d}t'w(|\mathbf{E}\left(t',r,\theta,z)|\right )\right ]\right),
\end{equation}
$I_p$ is the ionisation potential of the filling gas, $e$ is the electron charge, $m_e$ the electron mass, $n_0$ the initial neutral gas density, and $w(|\mathbf{E}\left(t,r,\theta,z)|\right)$ is the Perelomov, Popov, Terent'ev (PPT) ionisation rate~\cite{Perelomov1966,Ilkov1992}.

Our code runs in parallel; for this paper each simulation was run over 72 CPU cores. The radial integrals are adaptively evaluated to ensure convergence at each step.

The pump pulse initial conditions were always for a linearly polarised pulse in the fundamental HE$_{11}$ mode. At the output the polarisation extinction ratio was always larger than 300~dB (our numerical accuracy), indicating that no depolarisation occurs in ideal conditions.

We include the modes HE$_{1m}$ with $m$ typically up to 10, and both polarisation orientations for each mode. We have verified, through extensive convergence tests, that no other mode types and symmetries are excited, and that no energy is transferred to any higher order HE$_{1m}$ modes.
In fact minimal energy is transferred to modes higher than HE$_{11}$.
What is transferred primarily consists of energy at specific high-order mode phase-matched RDWs (see Fig.~\ref{fig:uv}c and d)~\cite{Tani2014}.

\section*{Materials and Correspondence}
Correspondence and requests for materials should be addressed to John C. Travers at j.travers@hw.ac.uk.

\section*{Data availability}
The data that support the plots within this paper and other findings of this study are available from the corresponding author upon reasonable request.

\section*{Code availability}
The computer code used in this study will be made available upon reasonable request to the corresponding author.

\section*{Author contributions}
JCT proposed this work, the initial theory, and the experimental design; he also performed the numerical simulations and drafted the manuscript. All authors contributed to the experimental implementation and refinement, the analysis and discussion of the results, and the editing of the manuscript.

\section*{Competing financial interests}
The authors declare that they have no competing financial interests.

\section*{Acknowledgements}
This work was funded by the European Research Council (ERC) under the European Union's Horizon 2020 research and innovation program: Starting Grant agreement HISOL, No. 679649. This work used EPCC’s Cirrus HPC Service (https://www.epcc.ed.ac.uk/cirrus).

\clearpage

\renewcommand{\thepage}{S\arabic{page}}  
\renewcommand{\thesection}{S\arabic{section}}   
\renewcommand{\thetable}{S\arabic{table}}   
\renewcommand{\thefigure}{S\arabic{figure}}
\renewcommand{\theequation}{S\arabic{equation}}
\setcounter{page}{1}
\setcounter{section}{0}
\setcounter{figure}{0}
\setcounter{equation}{0}

\begin{center}
    {\huge
        Supplementary material for}
    \par
    \vspace{0.5cm}
    {\huge
        High-energy pulse self-compression and ultraviolet generation through soliton dynamics in hollow capillary fibres}
    \par
    \vspace{1cm}
    {\large
        Published in Nature Photonics (2019)}
    \par
    \vspace{1cm}
    {\large
        John C. Travers, Teodora F. Grigorova, Christian Brahms, Federico Belli}
    \par
    \vspace{0.2cm}
    j.travers@hw.ac.uk
    \par
    \vspace{0.2cm}
    School of Engineering and Physical Sciences, Heriot-Watt University, Edinburgh, EH14 4AS, United Kingdom
    \vspace{1cm}
\end{center}

\section*{Supplementary Figures}

\begin{figure*}[h!tb]
    \centering
    \includegraphics{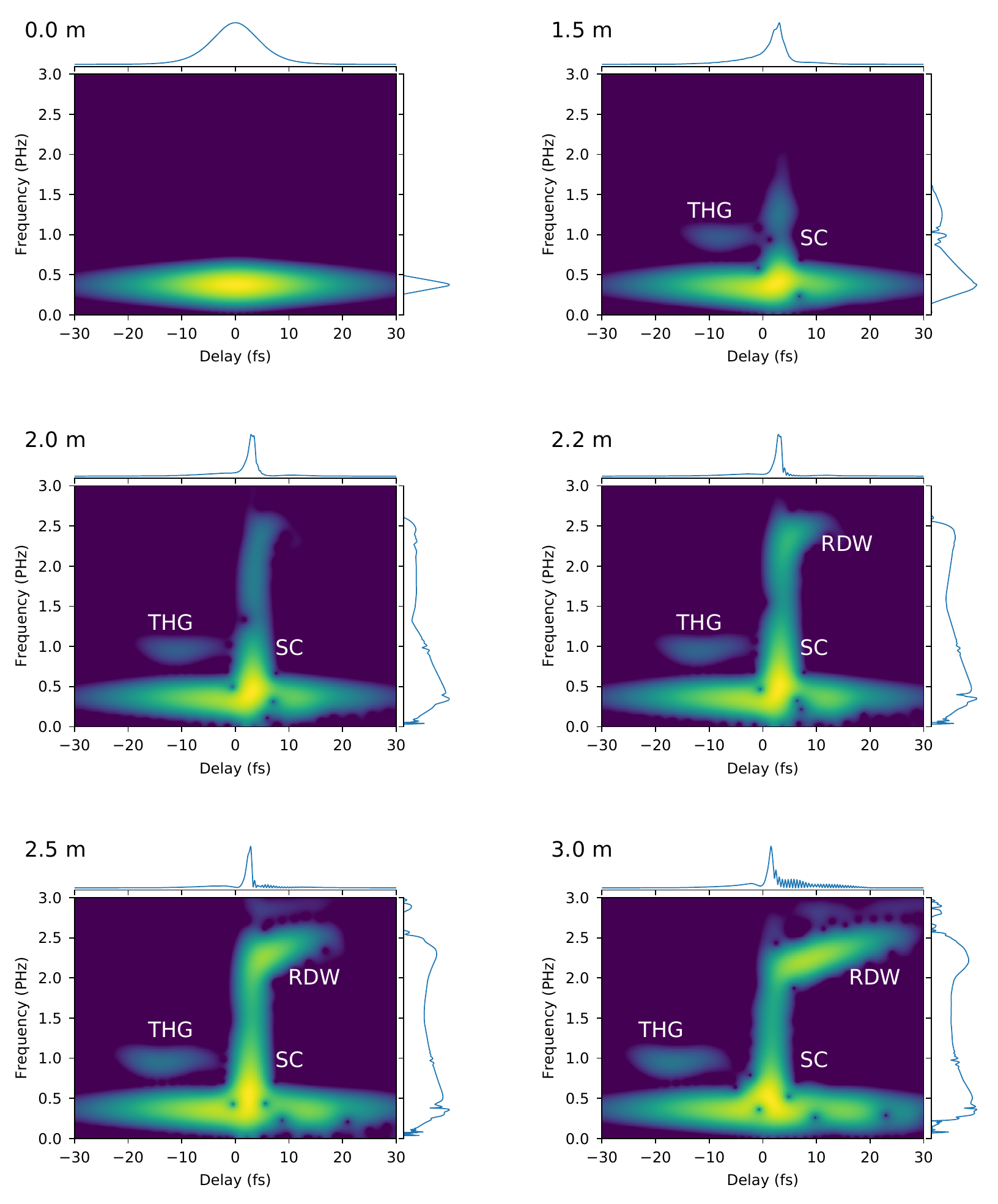}
    \caption{\textbf{Spectrograms of the evolution of the pulse in Fig.~\ref{fig:prop}b at the indicated positions in the HCF.} The colour scale is logarithmic over 50~dB and the gate pulse was a Gaussian with 2.6~fs FWHM. THG = third harmonic generation, SC = self-compression, RDW = resonant dispersive-wave.}
\end{figure*}

\begin{figure*}[h!tb]
    \centering
    \includegraphics{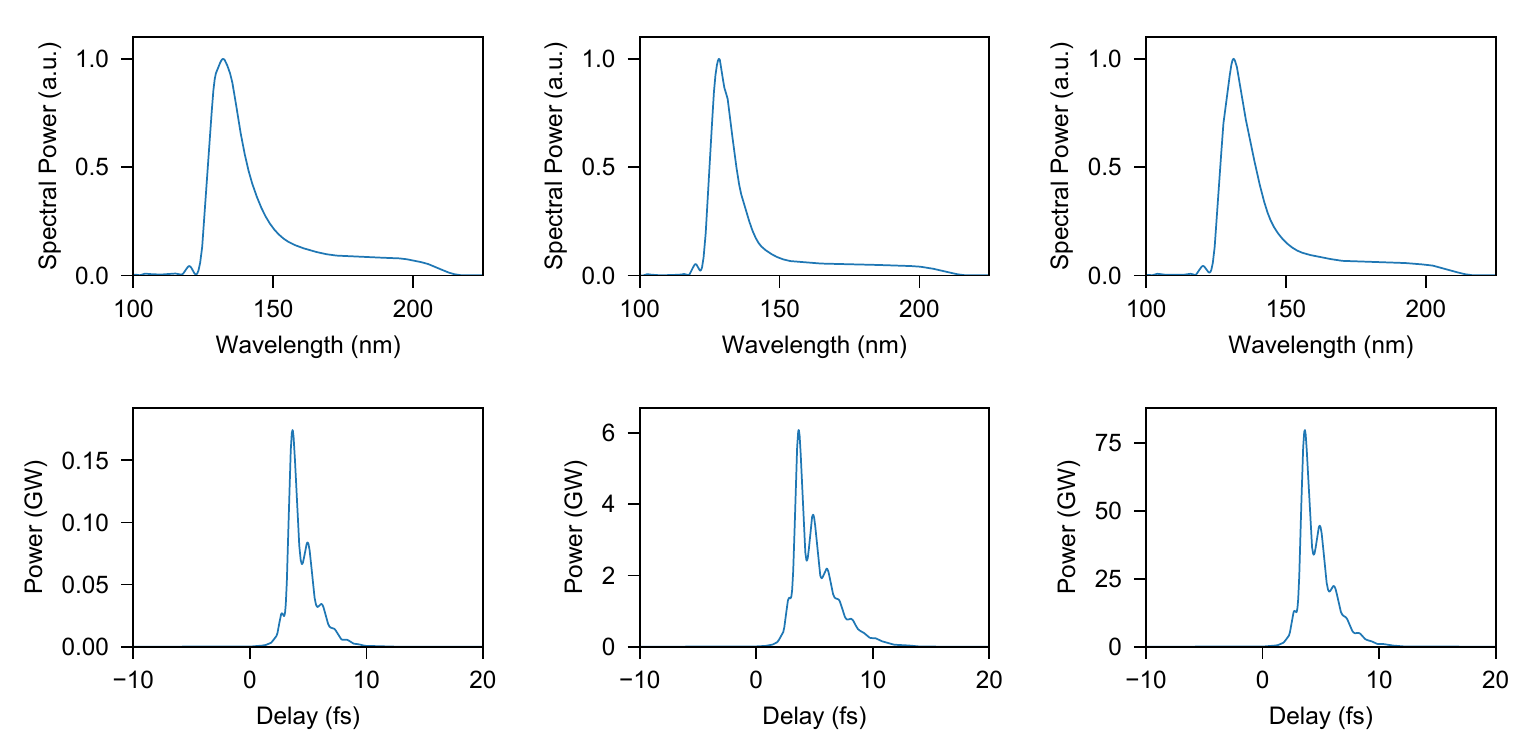}
    \caption{\textbf{Spectra and temporal pulse shapes and peak power of the RDW emission in Fig.~\ref{fig:prop}.} These are obtained by band-pass filtering the VUV part of the spectrum with a 100~nm wide super-Gaussian window centred at 150~nm, followed by inverse Fourier transforming. We selected the VUV pulses at the point at which they had highest peak power. Left to right correspond to Fig.~\ref{fig:prop}a, Fig.~\ref{fig:prop}b and Fig.~\ref{fig:prop}c. The temporal FWHM pulse durations are 1.8~fs and are not bandwidth limited. The bandwidth limited pulses supported by the RDW spectra have a duration of 0.8 fs.}
\end{figure*}

\begin{figure*}[h!tb]
    \centering
    \includegraphics{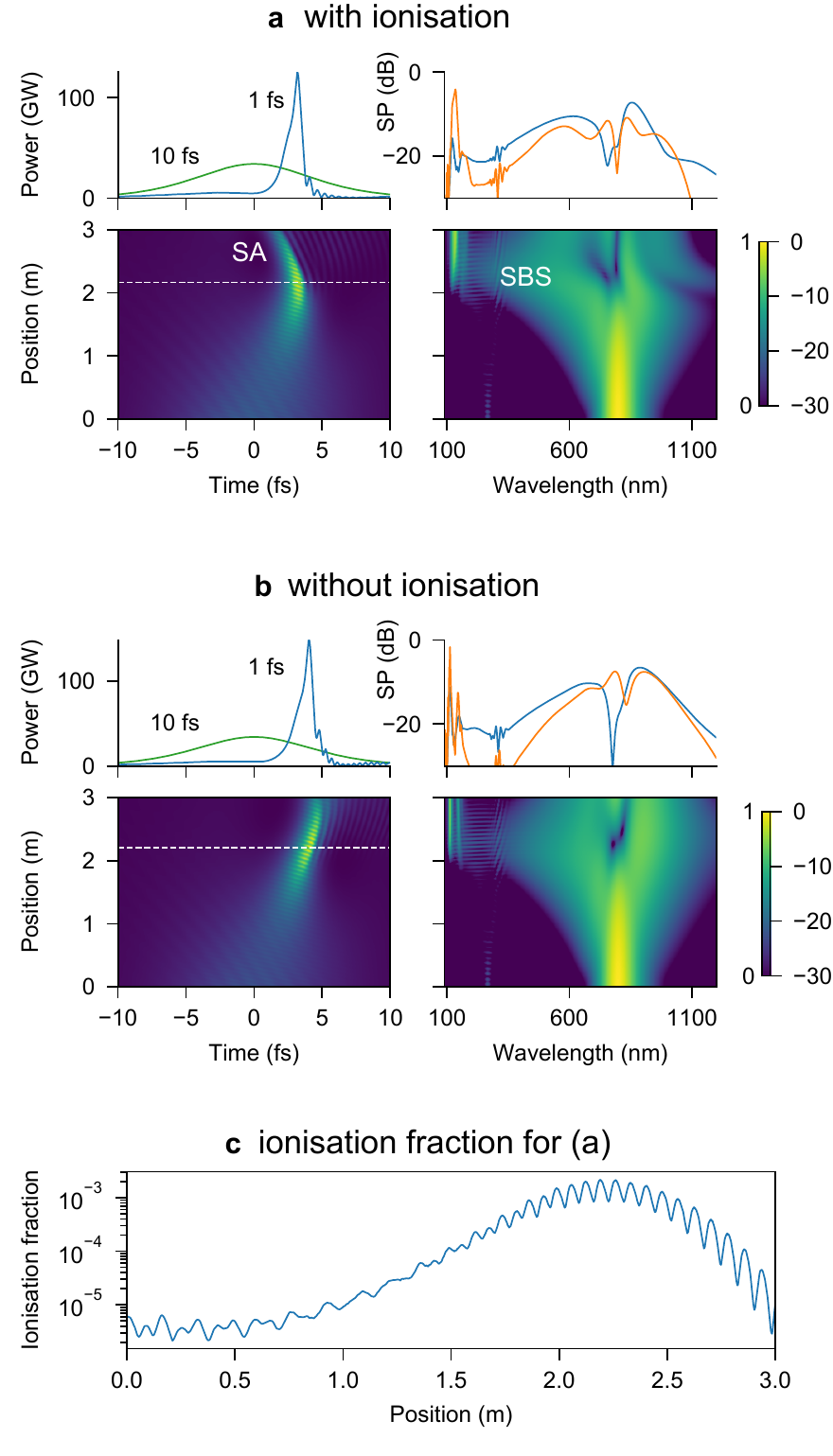}
    \caption{\textbf{Influence of ionisation and plasma dynamics.} Propagation plots for exactly the same parameters as Fig.~\ref{fig:prop}b (a) with full ionisation and plasma turned on; (b) with ionisation and plasma effects turned off. (c) Ionisation fraction as a function of position for (a). SA = soliton acceleration, SBS = soliton blue-shift.}
\end{figure*}

\begin{figure*}[h!tb]
    \centering
    \includegraphics[width=17.5cm]{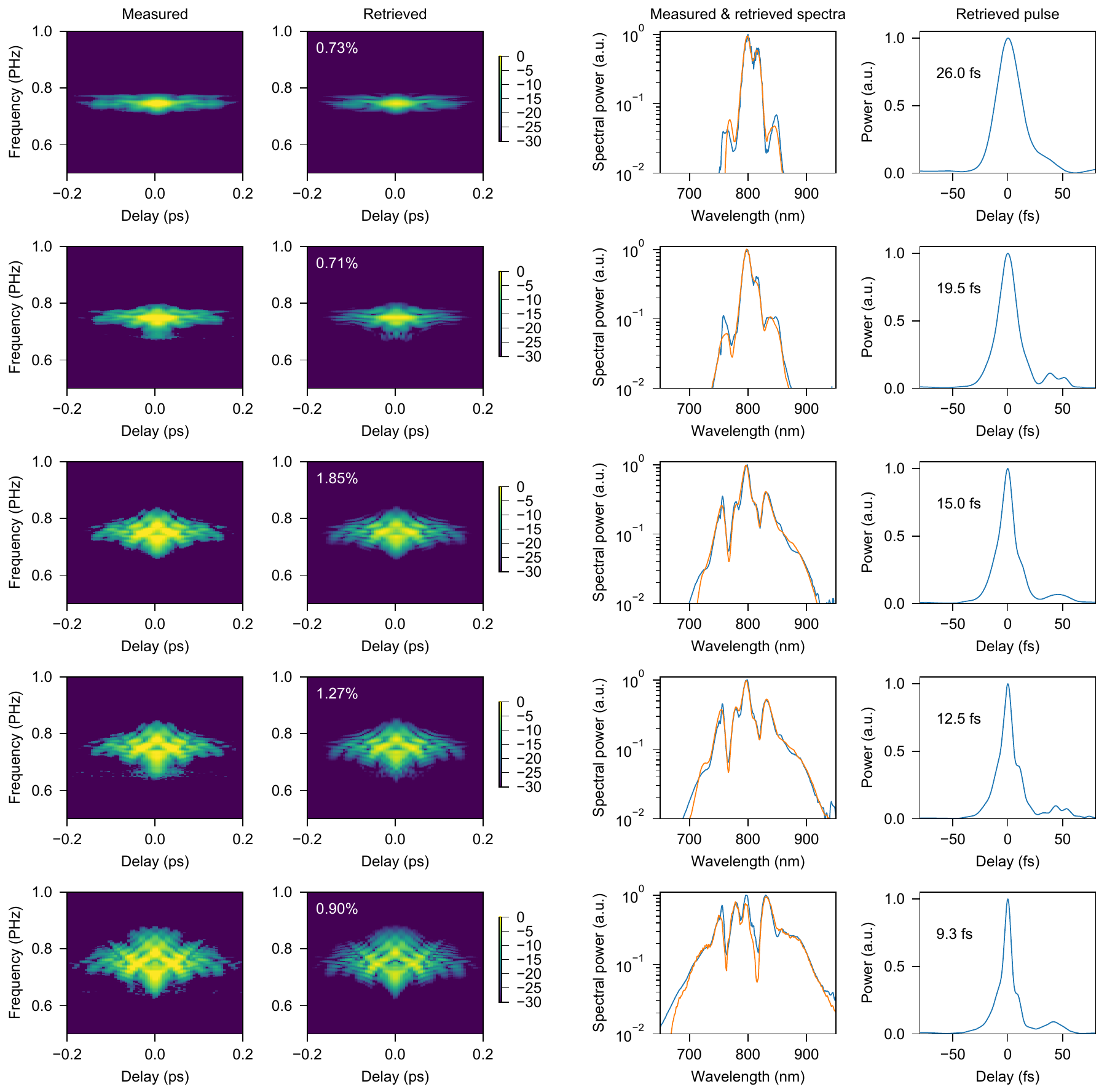}
    \caption{\textbf{Soliton self-compression pulse retrieval using SHG-FROG.} The rows from top to bottom correspond to the increasing energies in Fig.~\ref{fig:comp}a. First column: measured SHG-FROG traces; second column: retrieved SHG-FROG traces with the indicated FROG error; third column: measured and retrieved spectra; fourth column: retrieved pulses after back-propagation to the HCF output.}
\end{figure*}

\begin{figure*}[h!tb]
    \centering
    \includegraphics[width=17.5cm]{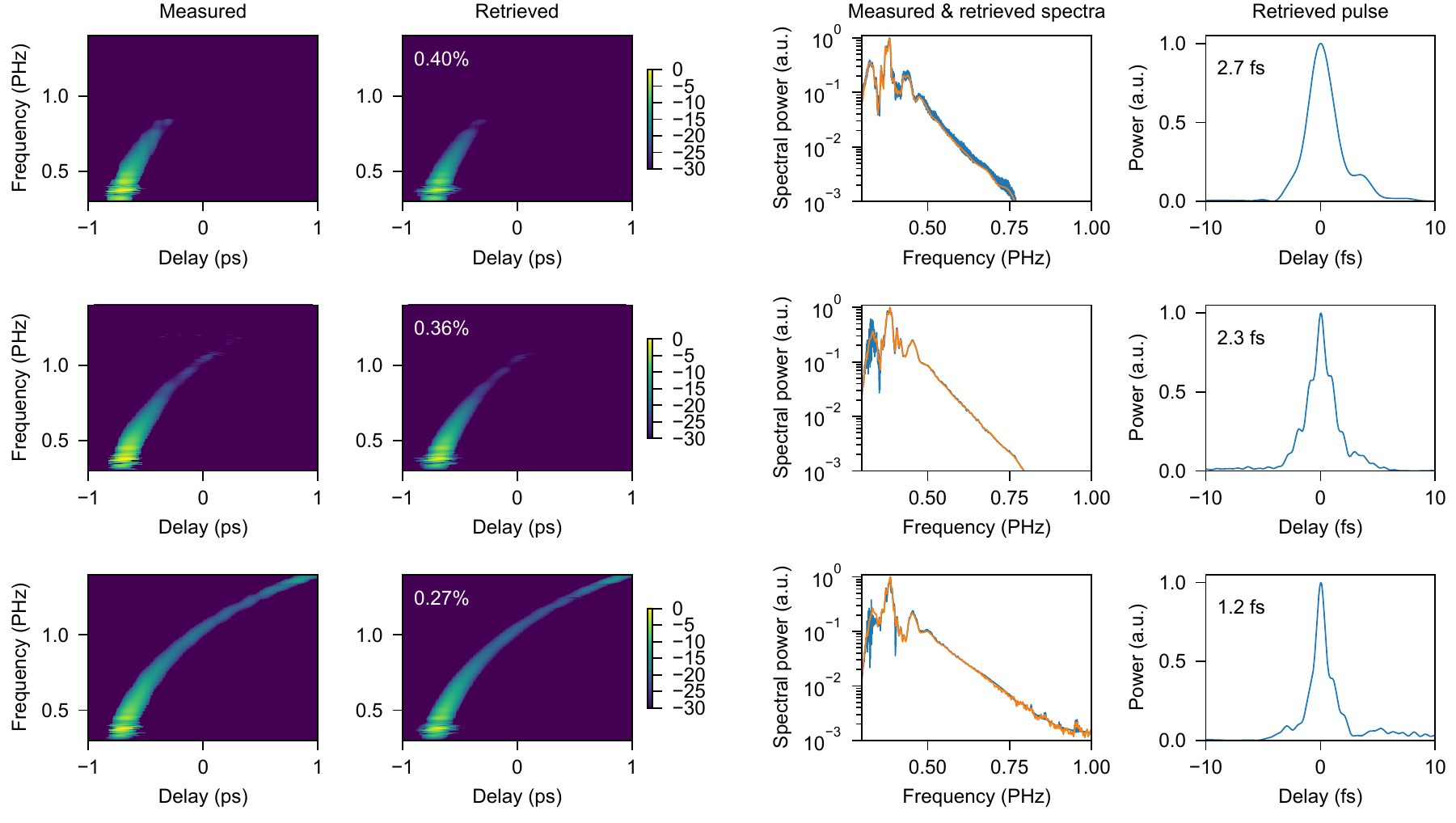}
    \caption{\textbf{Soliton self-compression pulse retrieval using PG-FROG.} The rows from top to bottom correspond to the increasing energies in Fig.~\ref{fig:comp}c. First column: measured PG-FROG traces; second column: retrieved PG-FROG traces with the indicated FROG error; third column: measured and retrieved spectra; fourth column: retrieved pulses after back-propagation to the HCF output. See the text surrounding Fig.~\ref{fig:comp}c for cautionary discussion about the short pulse durations measured here.}
\end{figure*}

\setlength{\tabcolsep}{2pt}
\renewcommand{\arraystretch}{1.0}
\begin{table}[h!tb]
    \centering
    \begin{ruledtabular}
        \begin{tabular}{@{}d{4.-1}d{3.0}d{3.0}d{1.1}d{3.0}d{2.1}d{2.1}@{}}
            \multicolumn{1}{c}{Pressure} &   \multicolumn{1}{c}{Energy\footnote{Estimated from low-energy coupling efficiency.}} &   \multicolumn{1}{c}{$\lambda_\mathrm{zd}$} & N &   \multicolumn{1}{c}{$\lambda_\mathrm{RDW}$\footnote{Calculated using the first moment of the RDW spectrum.}} & \multicolumn{1}{c}{RDW Energy} & \multicolumn{1}{c}{Efficiency\footnote{Estimated from low-energy coupling efficiency.}} \\
            \multicolumn{1}{c}{mb} &   \multicolumn{1}{c}{$\mu$J} &   \multicolumn{1}{c}{nm} &  &   \multicolumn{1}{c}{nm} & \multicolumn{1}{c}{$\mu J$} & \multicolumn{1}{c}{\%} \\
            \hline
            230 &               341 &                          334 & 1.8 &                           122 &                   1.8 &              0.5 \\
            276 &               341 &                          348 & 2.0 &                           134 &                   2.6 &              0.8 \\
            300 &               341 &                          355 & 2.1 &                           131 &                   2.9 &              0.9 \\
            350 &               341 &                          367 & 2.2 &                           134 &                   3.4 &              1.0 \\
            400 &               341 &                          379 & 2.4 &                           137 &                   4.1 &              1.2 \\
            452 &               341 &                          390 & 2.5 &                           141 &                   4.8 &              1.4 \\
            500 &               341 &                          399 & 2.7 &                           144 &                   5.5 &              1.6 \\
            550 &               341 &                          408 & 2.8 &                           147 &                   6.0 &              1.8 \\
            701 &               275 &                          431 & 2.9 &                           157 &                   8.9 &              3.2 \\
            801 &               256 &                          445 & 3.0 &                           164 &                  10.5 &              4.1 \\
            902 &               230 &                          457 & 3.0 &                           170 &                  11.5 &              5.0 \\
            1100 &               198 &                          479 & 3.1 &                           183 &                    &               \\
            1200 &               173 &                          489 & 3.1 &                           191 &                    &               \\
            1400 &               131 &                          508 & 2.9 &                           208 &                  14.9 &             11.4 \\
            1600 &               153 &                          524 & 3.4 &                           223 &                  16.2 &             10.6 \\
            1800 &                88 &                          539 & 2.8 &                           247 &                   8.2 &              9.3 \\
            2000 &                88 &                          553 & 3.0 &                           260 &                  10.3 &             11.7 \\
            2200 &                88 &                          565 & 3.2 &                           270 &                  12.9 &             14.7 \\
            2400 &                88 &                          577 & 3.4 &                           275 &                  13.1 &             14.9 \\
            2600 &                88 &                          589 & 3.6 &                           286 &                  13.0 &             14.8 \\
            3000 &                79 &                          609 & 3.8 &                           300 &                  10.4 &             13.2 \\
            3200 &                69 &                          619 & 3.7 &                           314 &                   9.6 &             13.9 \\
            3600 &                60 &                          637 & 3.8 &                           336 &                   6.4 &             10.7 \\
            4000 &                60 &                          653 & 4.1 &                           350 &                   6.3 &             10.5 \\
        \end{tabular}
    \end{ruledtabular}
    \caption{\textbf{Parameters for RDW tuning in Fig.~\ref{fig:uv}, with $a=125$~$\mu$m, 10~fs pump pulses and helium gas.}}
\end{table}

\begin{figure*}[h!tb]
    \centering
    \includegraphics[width=14cm]{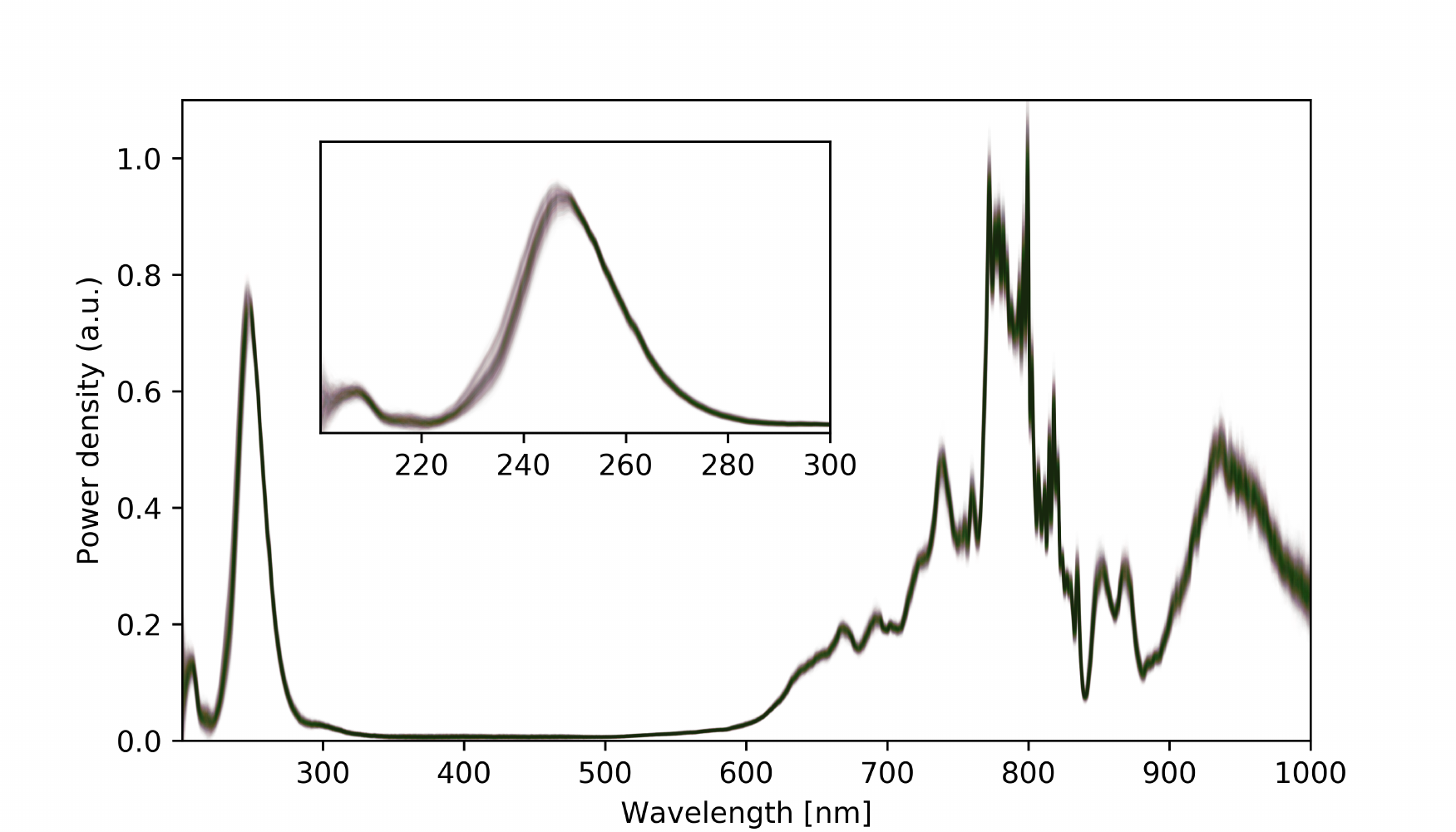}
    \caption{\textbf{Stability of RDW emission.} One thousand single-shot spectra of RDW emission at 252~nm in the 3~m HCF with $a=125$~$\mu$m pumped with 10~fs pump pulses. Each spectrum is slightly transparent, such that the opacity of the plot builds-up where the spectra overlap, providing a visual indication of the stability of the system. It is clear that the RDW emission is highly stable. The relative intensity noise (RIN) at the peak of the RDW of 252~nm is just 1.4\%, which is less than or comparable to the rest of the spectrum.}
\end{figure*}

\begin{figure*}[h!tb]
    \centering
    \includegraphics[width=14cm]{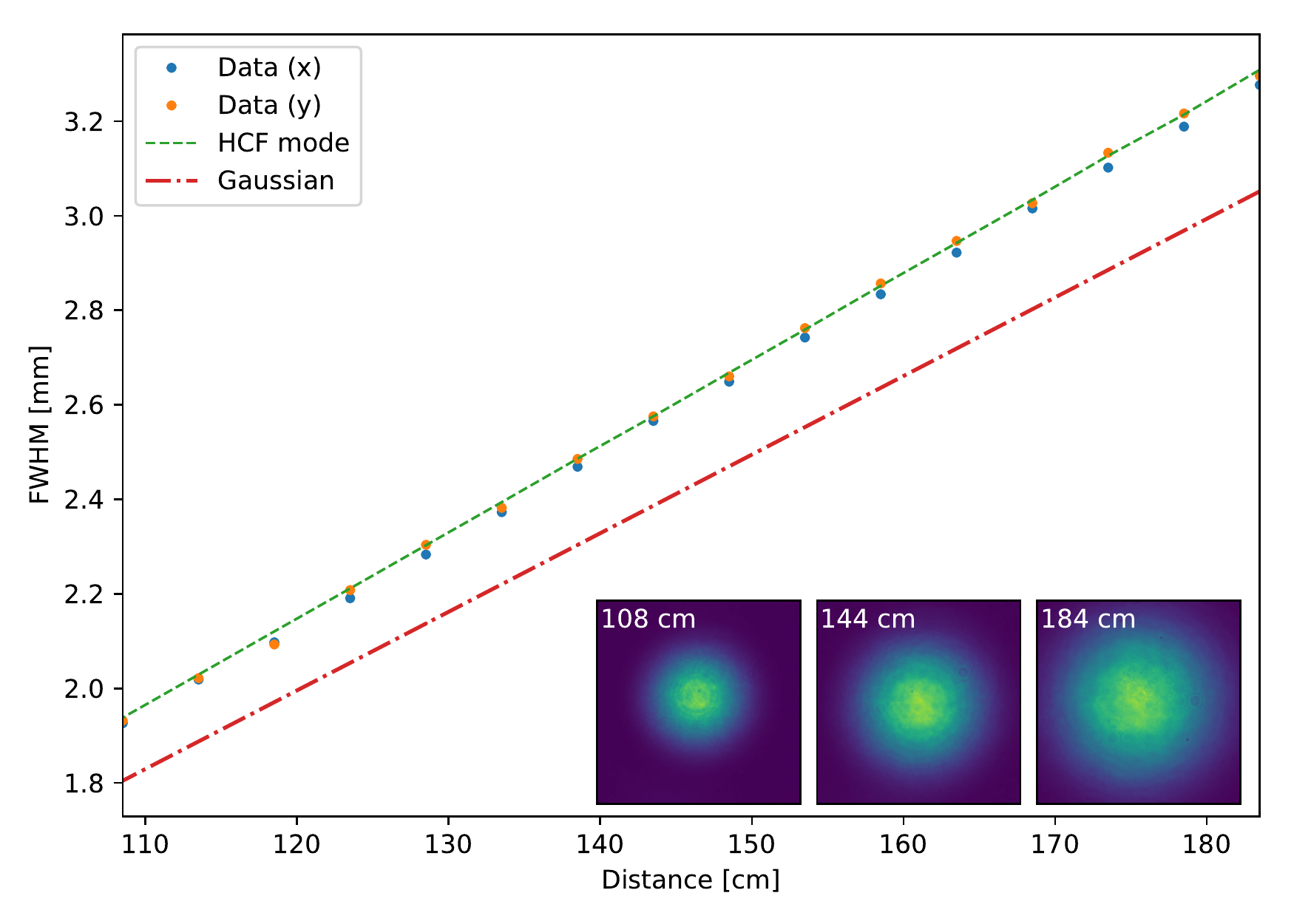}
    \caption{\textbf{The divergence of RDW emission.} The measured divergence of the RDW generated at 355~nm in the 3~m HCF with $a=125$~$\mu$m pumped with 10~fs pump pulses. The green dashed line indicates the theoretically expected divergence of the fundamental Bessel mode of the HCF obtained by numerically propagating it in free-space~\cite{Nagy2008}, the data points are our divergence measurements of the HCF output after passing through a 355~nm bandpass filter, and the red dashed line shows the Gaussian beam divergence for the same spot size. The insets show sample images of the beam at different distances from the HCF exit. The small features on the beam images are due to imperfections of the filtering and attenuation optics.}
\end{figure*}

\begin{figure*}[h!tb]
    \centering
    \includegraphics[width=17cm]{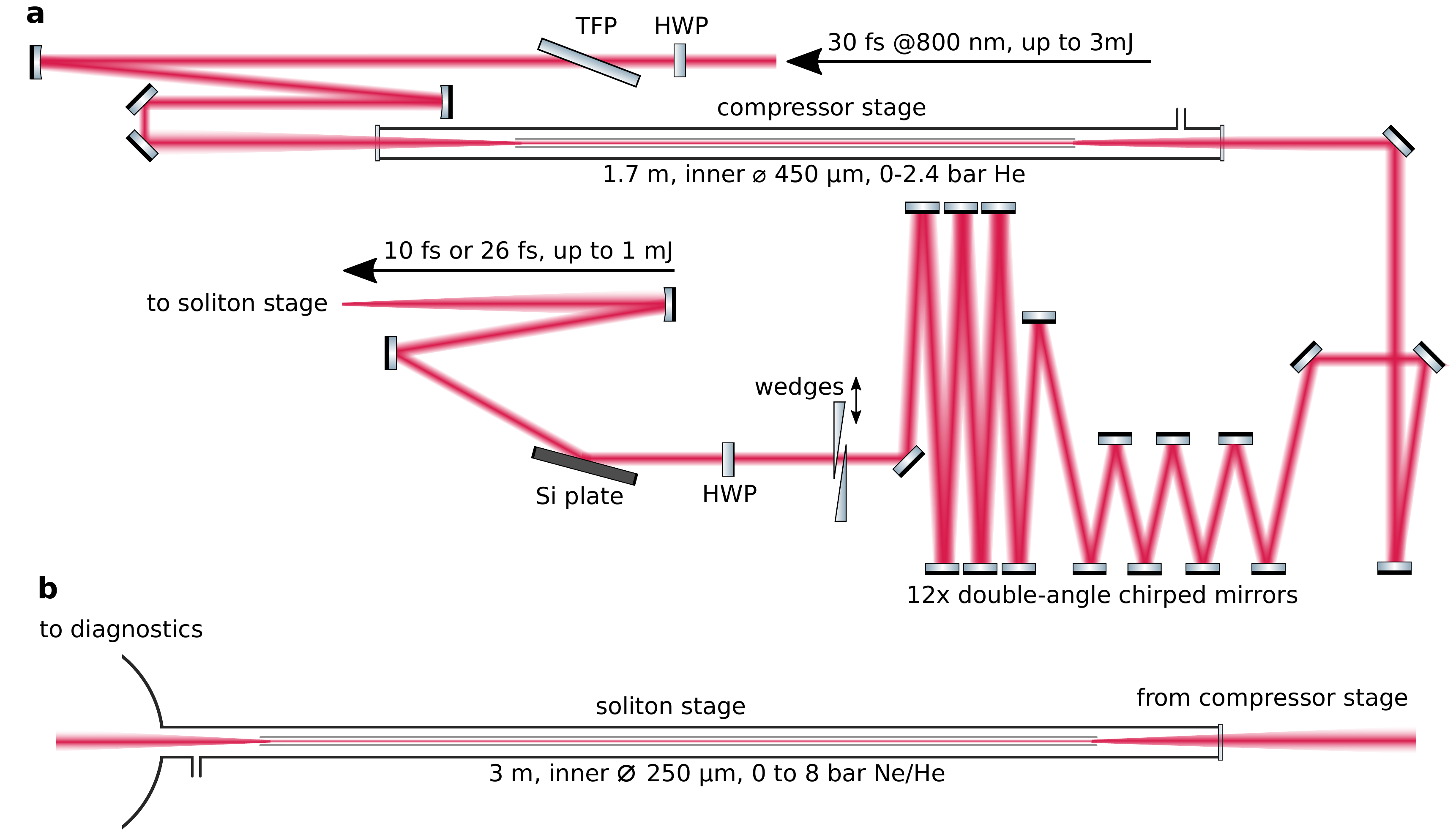}
    \caption{\textbf{Experimental setup.} (a) The compression stage. (b) The soliton stage. (TFP) Thin film polarizer, (HWP) Half-wave plate.}
\end{figure*}

\begin{figure*}[h!tb]
    \centering
    \includegraphics[width=9cm]{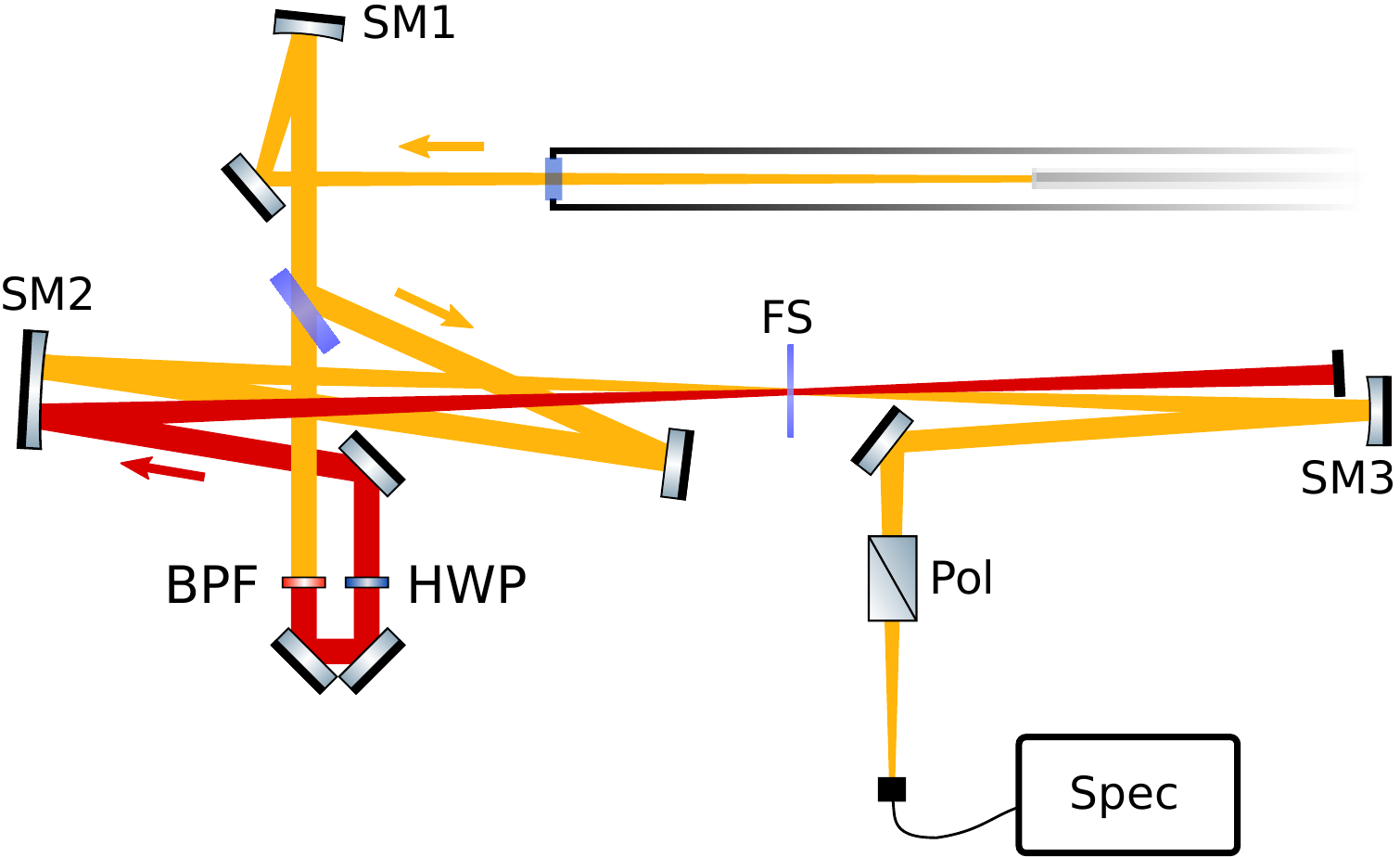}
    \caption{\textbf{Layout of the PG-FROG device.} After collimating the beam exiting the capillary using a spherical mirror (SM1), a purely s-polarised portion of the pulse is split off by Brewster-angle reflection off a wedged Magnesium Fluoride (MgF2) window (Thorlabs). The transmitted part is used to prepare the gate pulse by passing it through a bandpass filter (BPF) with a passband centred at 780~nm and 10~nm wide (Thorlabs) as well as a half-wave plate (HWP, Thorlabs) which rotates the polarisation of the beam by (45$^\circ$). A retro-reflecting pair of mirrors on a motorised delay stage (Physik Instrumente) provides the ability to scan the delay between the unknown pulse and the gate pulse. The two pulses are re-combined by focusing them into a 20~$\mu$m thick piece of UV fused silica (FS) using a spherical mirror (SM2). Subsequently, the polarisation-gating signal is isolated by a BBO Rochon-type polarizer (Pol, Edmund Optics) and collected by re-imaging the focus onto a fibre-coupled spectrometer (StellarNet).}
\end{figure*}

\renewcommand{\arraystretch}{1.2}
\begin{turnpage}
    \begin{table}
        \begin{ruledtabular}
            \begin{tabular}{@{}ld{4.1}d{2.1}ccd{3.2}d{3.1}d{3.0}d{2.2}c@{}}
                \multirow{3}{*}{Method} & \multicolumn{2}{c}{Pump parameters\footnote{We take the pump parameters at the fundamental wavelength of the driving laser, but after any pulse compression, shaping or coupling.}} & & \multicolumn{4}{c}{VUV parameters} &  & \multirow{3}{*}{Reference}\\
                \cline{2-3} \cline{5-8}
                & \multicolumn{1}{c}{Energy} & \multicolumn{1}{c}{Duration} && \multicolumn{1}{c}{Wavelength} & \multicolumn{1}{c}{Energy\footnote{We take the highest energy over the tuning range.}} & \multicolumn{1}{c}{Duration} & \multicolumn{1}{c}{Peak power\footnote{We take the highest estimated peak power over the tuning range.}} & \multicolumn{1}{c}{Efficiency\footnote{We take the highest estimated efficiency over the tuning range.}} & \\
                & \multicolumn{1}{c}{$\mu$J} & \multicolumn{1}{c}{fs} && \multicolumn{1}{c}{nm} & \multicolumn{1}{c}{$\mu$J} & \multicolumn{1}{c}{fs} & \multicolumn{1}{c}{GW} & \multicolumn{1}{c}{\%} & \\
                \hline
                RDW in HCF (demonstrated) & < 340.0 & 10.0 && continuous 110 to $>200$ & 13.00 & <2.0 & \sim 6 & 5.00 & This paper\\
                RDW in HCF (proposed) & 3000.0 & 10.0 && continuous 110 to $>200$ & 300.00 & <2.0 & \sim 80 &  20.00 & This paper\\
                RDW in HC-PCF & <5.0 & 35.0 && continuous 110 to $>200$ & 0.05 & & & 1.00 & \cite{Belli2015,Ermolov2015}\\
                Cascaded FWM in HCF & 100.0 & 25.0 && 160 or 200 & & >6 & & 0.01 & \cite{Misoguti2001}\\ 
                Noncollinear FWM in gas cell& 2800.0 & 42.0 && 160 & 2.5 & 43.0 & \sim 0.06 & 0.09 & \cite{Ghotbi2010}\\ 
                Collinear FWM in gas cell& 50000.0 & 100.0 && 89, 100, 114, 133, 160, 200 & 7.6 & >70.0 & \sim 0.1 & 0.02 & \cite{Shi2013}\\
                Collinear FWM in gas cell& 2461.0 & 50.0 && 79, 89, 133 & 6.4 & & & 0.26 & \cite{Zhou2014}\\
                FEL (DESY) & & & & continuous 95 to 105 & 100.0 & 50.0 & \sim 1 & & \cite{Ayvazyan2002}\\
                FEL (DCLS) current & & & & continuous 50 to 150 & 500.0 & 1500.0 & \sim 0.33 & & \cite{Chang2018}\\
                FEL (DCLS) proposed & & & & continuous 50 to 150 & 500.0 & 100.0 & \sim 5 & & \cite{Chang2018}\\
            \end{tabular}
        \end{ruledtabular}
        \caption{\textbf{Comparison of some schemes for generating ultrafast ($<100$~fs) pulses in the VUV spectral region (100~nm, to 200~nm).}}
    \end{table}
\end{turnpage}

\begin{figure*}[h!tb]
    \centering
    \includegraphics[width=9.5cm]{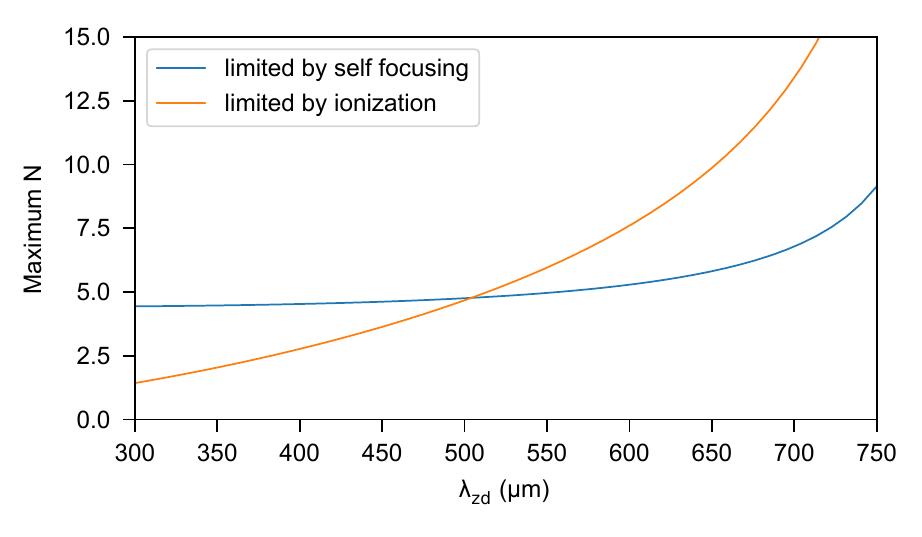}
    \caption{\textbf{The maximum allowed pump soliton order, limited by either self-focusing or ionisation, as a function of zero dispersion wavelength, for 10~fs pump pulses at 800~nm in helium.} These curves are independent of HCF core radius $a$, and directly proportional to the pump duration. See supplementary text for full details. From these curves we see that soliton self-compression and subsequent fission dynamics can be observed in HCF for zero dispersion wavelengths ranging from below 300~nm all the way to the pump wavelength.}
\end{figure*}

\clearpage

\section*{Supplementary Text}
\section{Derivation of soliton fission length scaling\label{sec:sup:scale}}
We are looking for scaling rules for the characteristic soliton fission length in terms of the core size $a$, the pulse duration $\tau_\mathrm{fw}$, and the soliton order $N$ or pump intensity $I_0$, while keeping the pump wavelength $\lambda$ and the zero dispersion wavelength $\lambda_\mathrm{zd}$ fixed (i.e. preserving the dispersion landscape experienced by the pump pulse).

The propagation constant (axial component of the wavevector) of the HE$_{nm}$ mode of a gas-filled hollow capillary fibre is given by~\cite{Marcatili1964}
\begin{equation}
\label{eqn:mcm}
\beta = \frac{\omega}{c}\sqrt{n_\mathrm{gas}^2-\frac{u_{nm}^2c^2}{a^2\omega^2}},
\end{equation}
where $\omega$ is the angular frequency, $c$ is the speed of light in vacuum, $n_\mathrm{gas}$ is the refractive index of the filling gas, $u_{nm}$ is the $m^\mathrm{th}$ zero of the Bessel function $J_{n-1}$, and $a$ is the core radius. Using $n_\mathrm{gas}^2=1+\rho_r\chi_e$, where $\chi_e$ is the frequency dependent susceptibility of the gas at some standard conditions, and $\rho_r$ is the gas density relative to those conditions, we can approximate Eq.~\ref{eqn:mcm} as
\begin{equation}
\label{eqn:mcma}
\beta \approx \frac{\omega}{c}\left(1+\frac{\rho_r\chi_e}{2} -\frac{u_{nm}^2c^2}{2a^2\omega^2}\right).
\end{equation}
For ideal gases, if we define $\chi_e$ at room temperature and atmospheric pressure, then $\rho_r$ approximately corresponds to the gas pressure in bar; this is reasonable for helium, but becomes less accurate for the heavier gases~\cite{Markos2017}. The values for $\chi_e$ can be taken from Sellmeier equations~\cite{Ermolov2015,Borzsonyi2008}.

The GVD is defined as $\beta_2=\partial^2\beta/\partial\omega^2$. For HCF we obtain, from Eq.~\ref{eqn:mcma}, in terms of wavelength, Eq.~\ref{eqn:b2} in the main text
\begin{equation}
\label{eqn:b2s}
\beta_2(\lambda, \rho_r, a) = \frac{\lambda^3}{4\pi c^2}\left(\rho_r\frac{\partial^2 \chi_e}{\partial \lambda^2}-\frac{u_{nm}^2}{2\pi^2a^2}\right).
\end{equation}
Therefore, to obtain a certain zero dispersion wavelength we tune the gas density to
\begin{equation}
\label{eqn:rhozd}
\rho_r(\lambda_\mathrm{zd},a)=\frac{u_{nm}^2}{2\pi^2a^2f(\lambda_\mathrm{zd})},
\end{equation}
where
\begin{equation}
f(\lambda') = \left.\frac{\partial^2 \chi_e}{\partial \lambda^2 }\right |_{\lambda=\lambda'}.
\end{equation}

If we fix $\lambda_\mathrm{zd}$, then the GVD at any other wavelength can be written (using Eq.~\ref{eqn:rhozd} and Eq.~\ref{eqn:b2s})
\begin{equation}
\label{eqn:b2lzd}
\beta_2(\lambda, \lambda_\mathrm{zd}, a) = \frac{\delta (\lambda, \lambda_\mathrm{zd})}{a^2},
\end{equation}
where
\begin{equation}
\label{eqn:delta}
\delta (\lambda, \lambda_\mathrm{zd}) = \frac{u_{nm}^2\lambda^3}{8\pi^3c^2}\left(\frac{f(\lambda)}{f(\lambda_\mathrm{zd})}-1\right).
\end{equation}
Importantly, the quantity $\delta(\lambda, \lambda_\mathrm{zd})$ does not depend on the gas pressure or any of the fibre parameters.

The dispersion length is defined as~\cite{Agrawal2007}
\begin{equation}
L_\mathrm{d}=\frac{\tau_0^2}{|\beta_2|},
\end{equation}
where $\tau_0\propto \tau_\mathrm{fw}$ is the natural pulse duration (for $\mathrm{sech}^2(\tau/\tau_0)$ pulses, $\tau_\mathrm{fw}=2\tau_0\ln (1+\sqrt{2})$). In a gas-filled, HCF, we find that the dispersion length is given by
\begin{equation}
L_\mathrm{d} = \frac{\tau_0^2a^2}{\left|\delta (\lambda, \lambda_\mathrm{zd})\right|} \approx \frac{\tau_\mathrm{fw}^2a^2}{3\left|\delta (\lambda, \lambda_\mathrm{zd})\right|}.
\end{equation}
As $L_\mathrm{fiss}=L_\mathrm{d}/N$, we obtain the first part of Eq.~\ref{eqn:lfiss}:
\begin{equation}
L_\mathrm{fiss}\approx \frac{\tau_\mathrm{fw}^2a^2}{3N\left|\delta (\lambda, \lambda_\mathrm{zd})\right|}.
\end{equation}

The nonlinear coefficient of a fibre mode is defined as~\cite{Agrawal2007} $\gamma=2\pi n_2/\lambda A_\mathrm{eff}$, where the nonlinear refractive index depends on the gas density as $n_2=n_2^0 \rho_r$, where $n_2^0$ is the nonlinear refractive index at the same standard conditions as $\chi_e$ above, and can be obtained from the literature~\cite{Lehmeier1985}. For the fundamental, HE$_{11}$, mode in HCF we can closely approximate the effective mode area as $A_\mathrm{eff}\approx 3a^2/2$. Therefore, we can define the nonlinear coefficient in the fundamental HE$_{11}$ mode of HCF, when the pressure is tuned for a given zero dispersion frequency, as 
\begin{equation}
\label{eqn:gamma}
\gamma^\mathrm{zd}=\frac{4\pi n_2^0\rho_r^\mathrm{zd}}{3\lambda a^2}=
\frac{2n_2^0u_{11}^2}{3\pi a^4\lambda f(\lambda_\mathrm{zd})} \propto \frac{1}{a^4}.
\end{equation}

The nonlinear length is $L_\mathrm{nl}=1/\gamma P_0$ where $P_0$ is the peak power. In a single fibre mode, the effective intensity is given by $I_0=P_0/A_\mathrm{eff}$, and so, in HCF with the gas density tuned for a specific $\lambda_\mathrm{zd}$, we have
\begin{equation}
L_\mathrm{nl}=\frac{2}{3 \gamma I_0 a^2} \propto \frac{a^2}{I_0}.
\end{equation}

The soliton fission length can then be written
\begin{equation}
L_\mathrm{fiss}=\frac{L_\mathrm{d}}{N}=\sqrt{L_\mathrm{d}L_\mathrm{nl}}\propto \frac{\tau_\mathrm{fw} a^2}{\sqrt{I_0}},
\end{equation}
which is the second part of Eq.~\ref{eqn:lfiss}.

These scaling rules have been carefully checked against exact numeric solutions.

\section{Maximum soliton order in hollow core fibres\label{sec:sup:maxN}}
There are limits to the maximum soliton order that can be used for soliton fission.
Firstly, to avoid modulational instability, we keep $N<15$~\cite{Dudley2006}.
Secondly, we must ensure significant self-focusing does not occur.
Thirdly, we must ensure that the gas is not excessively ionised, reducing pulse energy and causing spatial defocusing.

\subsection{Limits to the soliton order due to self-focusing}
The critical power for self-focusing is~\cite{Fibich2000}
\begin{equation}
P_\mathrm{cr}^\mathrm{sf}\approx \frac{3\lambda^2}{4\pi n_2^0\rho_r}.
\end{equation}

Taking the maximum peak power to be $P_0=P_\mathrm{cr}^\mathrm{sf}/S$---where we conservatively reduce $P_\mathrm{cr}^\mathrm{sf}$ by a safety factor $S$ as we want to avoid spatial effects during propagation; we take $S=10$ and have verified this is conservative with numerical simulations---we can determine the maximum soliton order through $N^2=\gamma P_0\tau_0^2/|\beta_2|$. For a fixed choice of $\lambda$ and $\lambda_\mathrm{zd}$ we use the expressions for $\beta_2$ and $\gamma$ provided in the previous section (Eqs.~\ref{eqn:b2lzd} and \ref{eqn:gamma}). We then obtain the upper limit of the soliton order due to self focusing to be
\begin{equation}
\label{eqn:Nsf} 
N_\mathrm{max}^\mathrm{sf}(\lambda, \lambda_\mathrm{zd}, \tau_0) =\left( \frac{\tau_0^2\lambda}{S\left|\delta (\lambda, \lambda_\mathrm{zd})\right|}\right)^{\frac{1}{2}}.
\end{equation}
Interestingly, $ N_\mathrm{max}^\mathrm{sf}$ is independent of the gas nonlinearity $n_2^0$ and the core size. It depends on $\lambda$ and $\lambda_\mathrm{zd}$, and is proportional to the pulse duration. 

Equation~\ref{eqn:Nsf} is plotted in Fig.~S10 for 10~fs pump pulses at 800~nm in helium. At low pressures (low $\lambda_\mathrm{zd}$) it is stable at around $N=5$, increasing as $\lambda_\mathrm{zd}$ approaches the pump wavelength. In the low pressure region ionisation dominates. This curve can be scaled proportionally with $\tau_0$.

\subsection{Limits to the soliton order due to ionisation}
While significant ionisation is tolerable, and can be useful for certain plasma-soliton dynamics~\cite{Holzer2011}, it becomes detrimental when too much pulse energy is lost and the plasma defocusing becomes too large.
As a guide to the maximum tolerable intensity, we take $I_\mathrm{th}$, the intensity threshold for barrier suppression ionisation~\cite{Ilkov1992}, which for helium is $I_\mathrm{th} \approx 1.5\times10^{15}$~W/cm$^2$.
This limits the maximum power $P_0$ to be $P_\mathrm{cr}^\mathrm{ion} =A_\mathrm{eff}I_\mathrm{th}= 3a^2I_\mathrm{th}/2$.
We also reduce this by a factor $S=10$ here, however, using our rigorous numerical simulations, we have verified that the resulting values of $P_\mathrm{cr}^\mathrm{ion}/S$ are conservative, and can readily be exceeded (and indeed often are in experiment).
Following the previous section, this leads to a maximum soliton order, limited by ionisation
\begin{equation}
\label{eqn:Nion} 
N_\mathrm{max}^\mathrm{ion}(\lambda, \lambda_\mathrm{zd}, \tau_0) =\left(\frac{\tau_0^2n_2^0I_\mathrm{th}u_{11}^2}{S\pi\lambda\left|\delta (\lambda, \lambda_\mathrm{zd})\right|f(\lambda_\mathrm{zd})}\right)^{1/2}.
\end{equation}
$N_\mathrm{max}^\mathrm{ion}$ is also independent of core size $a$, and is proportional to the pulse duration.

Equation~\ref{eqn:Nion} is plotted in Fig.~S10 for 10~fs pump pulses at 800~nm in helium. At low pressures (low $\lambda_\mathrm{zd}$) ionisation is the strictest limit on $N$. At higher pressures, as $\lambda_\mathrm{zd}$ approaches the pump wavelength, self-focusing becomes more dominant.

\section{Brightness\label{sec:sup:brightness}}
For comparing pulsed light sources we use the peak brightness, defined as the peak photon flux per unit cross-sectional source area per unit solid angle divergence. For a diffraction limited Gaussian beam the brightness in terms of peak power rather than photon flux can be shown to be
\begin{equation}
B_\mathrm{power}=\frac{P_0}{\lambda^2},
\end{equation}
where $P_0$ is the peak power and $\lambda$ is the wavelength. Converting this to photon flux gives
\begin{equation}
\label{eqn:sup:brightness}
B=\frac{P_0}{hc\lambda},
\end{equation}
where $h$ is Planck's constant and $c$ is the speed of light in vacuum.
Therefore, when comparing diffraction limited sources at the same wavelength, only the peak power is important for comparing peak brightness.

As RDW emission is perfectly diffraction limited (albeit not a Gaussian beam), as shown in Fig.~S7, we can use knowledge of Eq.~\ref{eqn:sup:brightness} to compare RDW emission to other sources, and in particular free electron lasers (FELs). Note however, that FELs are not perfectly diffraction limited, so such a comparison would actually overestimate the FEL brightness. Table~S2 shows a comparison of estimated peak powers for a variety of VUV generation schemes, indicating that RDW emission in HCF as described in this paper can provide peak brightness in the VUV region comparable to an FEL.

\subsection{Spectral Brightness}
It is common in the FEL community to use a definition of brightness that involves spectral purity, defined as the peak brightness within 0.1\% relative bandwidth. While most applications which require pulsed sources depend only on peak power and pulse duration, and not on spectral purity, it is interesting to make a comparison using standard quantities. That definition can be written as
\begin{equation}
B_\mathrm{spec}=\frac{P_0}{hc\lambda \frac{(\Delta \lambda/\lambda)}{0.001}}=\frac{10^{-3}}{hc}\frac{P_0}{\Delta \lambda},
\end{equation}
where $\Delta\lambda$ is the spectral width.

The relative bandwidth of the RDW emission in Fig.~\ref{fig:prop} and Fig.~S2 is 10\%, whereas the relative bandwidth of the FEL emission in~\cite{Ayvazyan2002} was 1\% (note that this is purely the result of the pulse duration being $> 10$ times longer). Taking the relative estimated peak powers of 6~GW and 1~GW for the RDW and FEL source respectively, we find that the brightness of the FEL source is approximately 2 times the RDW emission from the table-top source presented here. However, the scaled version of the source described here (Fig.~\ref{fig:prop}c) is predicted to have ten times higher peak power, and hence ten times higher brightness, while still being only 12~m long (table-top for those with a big table).

A more recent FEL source dedicated to the VUV region is at DCLS~\cite{Chang2018}. While it is predicted to reach pulse durations of 100~fs, it currently produces 1.5~ps pulses with $\sim0.33$~GW peak power in a relative bandwidth of 0.06\%. Therefore it currently has a peak brightness ten times lower than the above described sources.
\end{document}